\documentclass[preprint2]{aastex}
\usepackage{aastexug}

\shorttitle{Pomp\'eia et al.}
\shortauthors{Detailed analysis of bulge-like stars I. Stellar parameters, kinematics
and oxygen abundances}

\begin{document}

\input{epsf}

\title{Detailed Analysis of Nearby Bulge-like Dwarf Stars I. Stellar Parameters, Kinematics
and Oxygen Abundances}

\author{Luciana Pomp\'eia and Beatriz Barbuy}
\affil{Instituto Astron\^omico e Geof{\'i}sico, USP, 01060-970 S\~ao Paulo,
Brazil }
\email{pompeia@iagusp.usp.br, barbuy@iagusp.usp.br}

\and
\author{Michel Grenon}
\affil{Observatoire de Gen\`eve, Chemin des Maillettes 51, CH-1290 Sauverny, Switzerland}
\email{Michel.Grenon@obs.unige.ch}

\altaffiltext{1}{Based on observations carried out at the European Southern Observatory,
 La Silla}


\begin{abstract}

High resolution \'echelle spectra were obtained with the FEROS 
spectrograph at the 1.5m ESO telescope for 35 nearby bulge-like stars with metallicities 
in the range -0.8 $\leq$ [Fe/H] $\leq$ +0.4. Geneva photometry, astrometric data from Hipparcos 
and radial velocities from Coravel are available for these stars. From Hipparcos data it appears 
that the turnoff of this population indicates an age of 10-11 Gyr (Grenon 1999).

Detailed analysis of the sample stars is carried out and atmospheric 
parameters derived from spectroscopic and photometric determinations are presented. 
Oxygen abundances are derived based on the forbidden [O I] $\lambda$ 6300.3 {\rm \AA} line. 
The results show an oxygen overabundance pattern for most of the sample stars when compared to 
their disk counterparts.

\end{abstract}



\section{Introduction}

Stars in the solar neighborhood comprise a variety of stellar populations, with different 
kinematics, ages, chemical compositions and origins. The study of such populations 
can help understanding the formation and evolution of the Galaxy components.

With the aim of improving knowledge about these populations,  
an observational program of 7900 nearby stars was proposed for the Hipparcos mission, 
selected by M. Grenon, based on the NLTT proper motion catalogue (New Luyten Two Tenths 
Catalogue) (see Grenon 1999). Metallicities and temperatures were determined
from Geneva photometry  for 5500 of these stars.

Two subsamples of these stars, the MR (metal-rich) and the SMR 
(super metal-rich) stars, have been studied by Grenon  (1972, 1989, 1990, 1999, 2000), 
where MR are defined as having metallicities +0.08 $\leq$ [M/H] $\leq$ +0.20 and the SMR have 
[M/H] $>$ +0.20. Their isochronal ages revealed three groups with homogeneous
ages: a group of the Hyades generation, with $\sim$0.7 Gyr, an intermediate generation with 
3-4 Gyr, and a very old component of $\sim$ 10 Gyr.

The old SMR and MR population was found to be the flattest component of the solar neighborhood. 
The mean distance from the galactic plane for 219 of these SMR stars is only 0.16 kpc, which seems 
to be in contradiction with the classical $\sigma$W vs. age relation. 

The metallicity gradient of the galaxy indicates an inner disk or bulge origin for this component,
in which case a mechanism of radial migration with very little scatter in scale heights is required. 
Mergers and stochastic shocks with molecular clouds were proposed but they trigger little angular 
momentum change in stellar motions (Grenon 2000 and references therein). The most efficient mechanism 
suggested is the action of a galactic bar in the inner disk and bulge structures. Models by Fux (1997) 
predict that the bar perturbation induces motions of most stars towards the galactic center. 
Many of these stars merge to the bulge population acquiring its dynamics. A tiny fraction gains radial energy 
and drifts to higher galatocentric radius. They form a thin disk which expands outwards with their initial 
$\sigma$W nearly conserved (Grenon 1999, 2000, Raboud et al. 1998).

Sharing the same kinematical properties of the old MR and SMR samples, a collection of 612 
NLTT nearby stars is found. This sample is restricted to objects with highly eccentric orbits, 
pericentric distances R$_{p} <$ 5.5 kpc, and absolute vertical velocities of less than 20 kms$^{-1}$. 
The metallicity distribution of this flat component is identical to that seen in the bulge by 
McWilliam \& Rich (1994). Solar ratios are expected for disk stars while [$\alpha$/Fe] $\geq$ 0 
values are expected for bulge stars due to a fast chemical enrichment at early times, and a 
consequent dominant contribution by Type II SNe (Matteucci \& Brocato 1990, Matteucci et al. 1999, Moll\'a et al. 
2000). 
 
In the present work a subsample of the bulge-like stars with metallicities in the range 
-0.8 $\leq$ [Fe/H] $\leq$ +0.4 is studied. Detailed analyses are carried out and oxygen abundances 
are derived.  In Sect. 2 the kinematical properties of the sample are reported. In Sect. 3 the observations 
and reductions are described.  In Sect. 4 the effective temperatures, gravities and metallicities are derived. 
Oxygen abundances are presented in Sect. 5. The results are discussed in Sect. 6 and in Sect. 7 a summary is given.


\section{Kinematics, Observations and Reductions}

The present sample consists of stars with very eccentric orbits (e $>$ 0.25), maximum distances 
to the galactic plane $Z_{\rm max} < 1$ kpc, and pericentric distances R$_{p}$, as small as 2-3 kpc. 
Their V space velocities are between -50 to -130 km s$^{-1}$ while the U velocities are from -150 to 
+100 km s$^{-1}$. Basic kinematical properties are reported in Table 1. Proper motions are from the NLTT 
catalog and the orbital parameters R$_{p}$ and R$_{a}$ were deduced using a galactic potential 
model by Magnenat (1982 and private communication, 1984), assuming the following solar parameters: galatocentric 
distance R$_{\odot}$ = 8.0 kpc, circular velocity V$_{c}$ = 228.7 km s$^{-1}$ and local density of 
0.15 M$_{\odot}$/pc$^{3}$ (Grenon 1987a, 1987b). Heliocentric radial velocities were determined from a list 
of Ca I, Na I, Fe I and Mg I lines in the spectral range 6000 - 7600 {\rm \AA}.

\placetable{tb1}

The observations were carried out at the 1.52m telescope at ESO, La Silla, using the Fiber Fed Extended 
Range Optical Spectrograph (FEROS) (Kaufer et al. 2000). The total spectrum coverage 
is 356 nm - 920 nm, with a resolving power of 48,000. Two fibers, with entrance aperture of 2.7 
arcsec, recorded simultaneously star light and sky background. The detector is a back-illuminated CCD
with 2048 $\times$ 4096 pixels of 15 $\mu$m size.

\par Sample stars were observed on five nights, September 22-26, 1999. Using a special package for 
reductions (DRS) of FEROS data, in MIDAS environment, the data 
reduction proceeded with subtraction of bias and scattered light in the CCD, orders extraction, flat 
fielding, and wavelength calibration with a ThAr calibration frame. 

\par Table 2 gives basic data for the sample stars: a short designation given in the present work, 
identification, V magnitude from Hipparcos, spectral type, distance (pc), reddening E(B-V), 
M$_{\rm v}$ (dereddened values), bolometric correction BC, stellar mass 
($\rm M_{\ast}$/$\rm M_{\odot}$), and the signal to noise ratio S/N. The S/N ratio was calculated and 
averaged for five regions free from lines. The reddening was derived by using the code Extinct from 
Hakkila et al. (1997) and using the option of extinction by Arenou et al. (1992). Bolometric corrections 
were inferred from the theoretical calibrations of Lejeune, Cuisinier \& Buser (1997). Stellar masses 
were derived from isochrones of Vandenberg (1985) and  VandenBerg \& Laskarides (1987).

 \placetable{tb2}
 
 
\section{Atmospheric Parameters}

\subsection{Temperatures}

\subsubsection{Photometry}

Geneva photometry was used to estimate the effective temperatures T$^{\rm Gen}$, by applying the 
calibration given in Grenon (1982). The derived values are given in Table 3. 

The effective temperatures were also calculated from Stromgren colors given in Hauck \& Mermilliod 
(1998). The Stromgren colors were transformed to temperatures T$^{\rm Strom}$ according to relation 
(9) given  in Alonso et al. (1999). Stromgren colors were then dereddened assuming the relation of 
E(b-y) = 0.13 E(B-V) given in Crawford \& Mandwewala (1976).

Alonso et al. (1996) and Blackwell et al. (1990) used the InfraRed Flux Method (IRFM) to derive the 
temperatures of a sample of F0-K5 dwarfs, among which 3 stars of the present sample. In Table 3 are 
reported the resulting temperatures T$^{\rm Gen}$, T$^{\rm Strom}$ and T$^{\rm IRFM}$ for the sample 
stars.

\subsubsection{H$\alpha$ profiles}

Spectroscopic temperatures were derived using H$\alpha$ profiles. These profiles were computed
with MARCS model atmospheres (Gustafsson et al. 1975) and a revised version of the code HYDRO by
Praderie (1967). The temperatures which better reproduce the H$\alpha$ wings profiles
were chosen. The continuum level was set using continuum regions on the red and blue 
sides of the Balmer feature. The reliability of temperatures derived from hydrogen lines profiles 
was discussed by Fuhrmann et al. (1993, 1994), by comparing temperatures determined from Balmer 
lines with those from photometric data. They found a smaller scatter among temperatures derived 
from Balmer lines than among those derived from the photometric indices $b-y$ and $V-K$.

In Fig. 1 we show the fit of the computed H$\alpha$ wings to the observed profile of HD 179764.
Synthetic spectra were computed for T$_{\rm eff}$  = 5350 K, 5450 K and 5450 K. In Table 3 
T$^{\rm H\alpha}$ values are given.

\begin{figure}
\label{fg1}
\epsscale{1.15}
\plotone{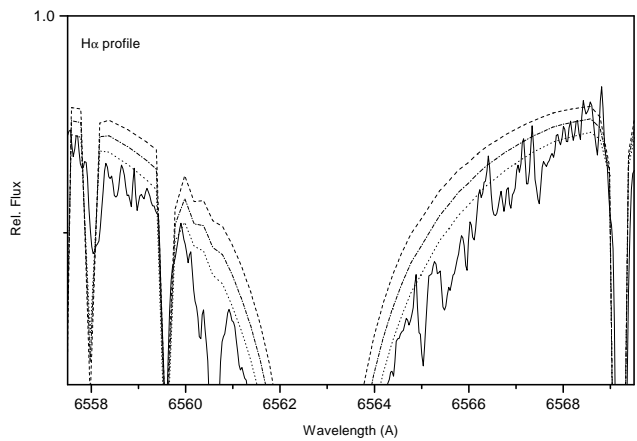}
\caption{Observed $H\alpha$ profile (solid line) of HD 179764, and synthetic spectra for 
 T$_{\rm eff}$  = 5550 K (dotted line), T$_{\rm eff}$  = 5350 K (dashed line), and the best fit for 
T$_{\rm eff}$  = 5450 K (dash-dotted line)}.
\end{figure}

\placefigure{fg1}

The mean differences of T$^{\rm H\alpha}$ relative to T$^{\rm Gen}$ and T$^{\rm Strom}$ are
$\Delta$(T$^{\rm H\alpha}$ $-$ T$^{\rm Gen}$) = 49 K with a standard deviation $\sigma$ = 42 K, 
and $\Delta$(T$^{\rm H\alpha}$ $-$ T$^{\rm Strom}$) = 69 K with $\sigma$ = 54 K. 
Note that Fuhrmann et al. (1994) calculated the temperature of HD 143016 using H$\rm \beta$ profiles 
computed with models scaled to empirical solar models. They found T$_{\rm eff}$  = 5650 K for this star, 
in good agreement with our derived value of T$_{\rm eff}$ = 5575 K.

Model atmospheres were checked for ten of our stars using two other grids of models, OSMARCS by
Edvardsson et al. (1993) and ATLAS by Kur\'ucz (1993). We also calculated temperatures with
H$\beta$ profiles T$^{\rm H\beta}$, using MARCS models. The resulting  T$^{\rm Kur}$ for Kur\'ucz (1993)
models, T$^{Edv}$ for Edvardsson et al. (1993) models, and T$^{\rm H\beta}$ are shown in Table 4
(the spectrum of CD-40 15036 has a bump in the H$\beta$ region, and T$^{\rm H\beta}$ was not derived
for this star). There is good agreement among the different temperature indicators and we adopted 
T$^{\rm H\alpha}$ as final temperature values.

\placetable{tb3}

\placetable{tb4}


\subsection{Gravities}

\par Trigonometric gravities were derived using Hipparcos parallaxes through the classical
formula:

\begin{displaymath}
log g/g_{\odot}= {{M_{\ast}\over {M_{\odot}}} + 4log {T_{\ast}\over T_{\odot}}+ 
0.4V + 0.4 BC + 2log{\pi} +0.12}
\end{displaymath}

\noindent

Spectroscopic gravities were calculated by requiring that \ion{Fe}{1} and \ion{Fe}{2} yield 
the same iron abundance. The line lists were selected from Castro et al. (1997), Nave et al. (1994) 
and from the NIST Atomic Spectra Database (\rm phy\-sics\-.nist\-.gov\-/cgi\--bin\-/AtData\-/main-asd).
Using the Atlas of the Solar Photosphere (Wallace, Hinkle and Livingston 1998) a check of possible 
blends was performed and blended lines were discarded. The line lists of \ion{Fe}{1} and \ion{Fe}{2} are 
given in Tables 5 and 6 respectively. The log gf values are from NIST and the adopted solar abundances 
are from Grevesse et al. (1996).

The results for log $g$ from Hipparcos parallaxes and ionization equilibrium are given in Table 11. 
The average difference between the two methods is of 0.15 dex. Discrepancies between these methods 
have been pointed out by Edvardsson (1988), Nissen et al. (1997), Fuhrmann et al. (1997) 
and Allende Prieto et al. (1999, hereafter AGLG99) in the sense that trigonometric gravities are higher 
than spectroscopic ones. AGLG99 studied the trigonometric and spectroscopic gravities available in the 
literature, revealing that the absolute differences are small for stars in the metallicity range 
$-1.0 < {\rm [Fe/H]} < 0.3$, with a mean $<$ 0.10 dex. For the present sample a mean difference 
$\Delta$(log g$^{\rm Spec}$ - log g$^{\rm Hip}$)$=$ 0.19 dex is found with a standard 
deviation $\sigma = 0.14$, in agreement with the AGLG99 results. We have adopted the 
spectroscopic gravities hereafter.

\placetable{tb5}

\placetable{tb6}


\subsection{Metallicities and Microturbulent Velocities}

The metallicities [Fe/H] and microturbulent velocities $\xi_{t}$ were derived using curves 
of growth of \ion{Fe}{1} and \ion{Fe}{2}. The curves were built with the code RENOIR by M. Spite.
Equivalent widths were measured using IRAF. The S/N of the spectra is high (Table 2) and the continua 
are well-defined.

 Curves of growth for \ion{Fe}{1} and \ion{Fe}{2} of HD 179764 are given in Fig. 2. Equivalent 
widths of \ion{Fe}{1} are given in Tables 7 and 8 and those of \ion{Fe}{2} in Tables 9 
and 10. Figs. 2a, b show the curves of growth of \ion{Fe}{1} and \ion{Fe}{2} for HD 179764. 
Metallicities inferred from Geneva photometry are also shown. In Table 11 the resulting values of 
[Fe/H] and $\xi_{t}$ are reported. 

\placetable{tb7}
\placetable{tb8}
\placetable{tb9}
\placetable{tb10}

Small differences are found between photometric and spectroscopic metallicities: the mean 
value of the difference is 0.16 dex with a standard deviation of 0.15 dex. The final adopted values 
[Fe/H]$^{Spec}$ are given in Table 11.  
 
 \placetable{tb11}
 
 
\section{Oxygen Abundances}

Oxygen abundances were derived by comparing the observed [O I] 6300.3 {\AA} line to synthetic spectra. 
The log gf value was adopted from Castro et al. (1997). The spectrum synthesis code is described 
in Barbuy (1988) and Barbuy \& Erdelyi-Mendes (1989). This line is insensitive to NLTE effects (Kiselman 2001). 
Telluric lines are removed by subtracting the spectrum of hot and rapidly rotating stars. 
In Figs. 3a,b we show the fit of the [O I] line for HD 149256 and HD 148530. In Table 12 the 
resulting oxygen abundances are given. Spectrum synthesis analysis to infer other $\alpha$ 
and heavy elements abundances is in progress and will be presented elsewhere.

\placefigure{fg3}

\placetable{tb12}

\subsection{Abundance Errors}
The abundances derived from the [O I] line are not very sensitive to temperature. A variation 
of $\Delta$T$_{\rm eff}$ = 100 K induces $\Delta$[O/Fe] $\approx$ 0.05 dex. Oxygen abundances 
are also almost insensitive to microturbulent velocity and [Fe/H] uncertainties, 
$\Delta$$\xi_{t}$ = 0.2 kms$^{-1}$ and $\Delta$[Fe/H] = 0.2 dex will result in 
$\Delta$[O/Fe] $\approx$ 0.05 and 0.02  respectively. [O/Fe] values are somewhat  more 
sensitive to the gravity choice: $\Delta$log g = 0.2  gives $\Delta$[O/Fe] = 0.1, but the 
errors in gravities are smaller than 0.2 dex, as shown by the differences between spectroscopic 
and trigonometric values.

In order to test how changes in atmospheric parameters correlate with each other and
with the derived oxygen abundance we have applied changes in the temperature of a few stars to values
$\pm$ 100 K the inferred one. For the lower temperatures, the ionization balance gives average 
gravity and metallicity values 0.2 dex and 0.06 dex below the derived ones, respectively, and the resulting 
[O/Fe] ratio changes by 0.06 dex. For the higher temperatures, the ionization balance gives average 
gravity and metallicity values 0.15 dex above and 0.07 dex below the derived ones, respectively, 
and the resulting [O/Fe] ratio changes by 0.12 dex.


\section{Discussion}

In Fig. 4 the [O/Fe] ratio is plotted against [Fe/H] for the present sample, including
also SMR stars studied by Barbuy \& Grenon (1990) and a sample of disk stars by Nissen \& 
Edvardsson (1992). The oxygen abundances of both samples were determined from the 
[O I] 6300 {\AA} line. Theoretical curves of inside-out models by Matteucci et al. (1999) 
for the bulge (solid line), and by Chiappini et al. (2001) for the solar neighborhood (dotted 
line), are also plotted.

\placefigure{fg4}

We found a higher [O/Fe] ratio for most of our stars when compared 
to their disk counterparts, although some of them show a disk-like pattern. 
Barbuy \& Grenon (1990) also found an enhanced oxygen abundance
for a SMR sample. A possible interpretation to such overabundance is that 
the flat component stars originate in a region of the Galaxy where the chemical 
evolution occurred faster than in solar neighborhood.

It is also apparent from Fig. 4 that the [O/Fe] ratios are below the predicted theoretical bulge
curve. In this model (Matteucci et al. 1999) a fast colapse of 10$^{8}$ yr is assumed. 
Nevertheless, Maciel (2001), based on oxygen abundances of bulge planetary nebulae, suggested 
that the theoretical [O/Fe] vs. [Fe/H] curve from Matteucci et al. (1999) overestimates the 
[O/Fe] ratio by 0.3 to 0.5 dex, claiming that the discrepancy between the bulge theoretical 
curve and the solar neighborhood curve is smaller.     

 
 \section {Summary}
 
 High resolution spectra of stars with -0.8 $\leq$ [Fe/H] $\leq$ +0.4 and isochronal ages of $\sim$ 10 Gyr 
were obtained with the FEROS spectrograph. The sample stars have kinematical properties typical 
of a bulge or inner disk origin.

The main results from the present work are:

(1) Temperatures derived from photometric data (Geneva and Stromgren), 
IRFM, and H$\alpha$ profiles show small discrepancies ($<$ 100 K) for most of the stars, 
with mean differences of $\sim$ 70 K.

(2) A good agreement between spectroscopic (ionization equilibrium) and 
trigonometric gravities is also found, with a mean difference of 0.19 dex.

(3) A mean difference of 0.16 dex is found for the derived metallicities from Geneva 
photometry and from curves of growth of \ion{Fe}{1} and \ion{Fe}{2}.

(4) The curves for [O/Fe] vs. [Fe/H] show that most of our sample stars are overabundant in oxygen
with respect to disk stars with the same metallicity.

\smallskip$Acknowledgments$  

L.P. thanks J. Melendez for very useful discussions. We thank C. Chiappini, 
F. Matteucci and D. Romano for sending the theoretical curves of oxygen chemical enrichment. 
L. P. acknowledges the FAPESP PhD fellowship n$^{\rm \circ}$ 98/00014-0. We acknowledge FAPESP project 
n$^{\rm \circ}$ 1998/10138-8.



\begin{deluxetable}{crrrrrrr}
\label{tb1}
\tabletypesize{\scriptsize}
\tablecaption{Kinematical Data}
\tablewidth{0pt}
\tablehead{
\colhead{Name} & \colhead{U} & \colhead{V} & \colhead{W}
& \colhead{V$_{\rm Helio}$} & \colhead{Rp} & \colhead{Ra}
}
\startdata
HD 143016  &  -3.2 & -88.6  &  26.3  &   9.16(0.18) & 3.80 & 8.02 \\
HD 143102  & 52.9  & -92.5  & -19.6  &   7.44(0.11) & 3.58 & 8.23  \\
HD 148530  &-80.8  & -79.9  & -34.1  &  25.27(0.14) & 3.83 & 9.15 \\
HD 149256  &-7.0   & -98.4  & -48.1  &  25.10(0.07) & 3.40 & 8.04 \\
HD 152391  & -85.2 & -110.1 & 8.8    &  45.19(0.24) & 2.78 & 9.01 \\
HDE 326583  & -45.0 & -100.4 & -39.4  &  74.42(0.27) & 3.25 & 8.36 \\
HD 175617  &-121.3 &  -72.9 & -28.7  &  71.00(0.17) & 3.79 & 10.56 \\
HD 178737  & -8.3  & -92.4  &  64.7  &  34.59(0.17) & 3.64 & 8.04  \\
HD 179764  & 21.2  & -106.3 &   6.1  & -66.40(0.16) & 3.11 & 8.01  \\
HD 181234  & 5.3   & -91.7  &  1.8   & -46.77(0.13) & 3.68 & 8.00  \\
HD 184846  & -5.1  &-123.7  & -42.6  &  91.22(0.13) & 2.48 & 8.02  \\
BD-176035  & 35.0  &-82.9   &  10.6  & -65.85(0.12) & 4.02 & 8.09   \\
HD 198245  & 4.4   &-128.2  &  21.6  & -20.63(0.30) & 2.34 & 8.00   \\
HD 201237  &-85.9  &-79.0   &  -0.8  &  30.17(0.05) & 3.83 & 9.29  \\
HD 211276  &-20.7  &-105.9  & -36.6  & -24.30(0.17) & 3.10 & 8.11  \\
HD 211532  & 73.1  &-87.4   &  49.9  &-111.24(0.20) & 3.70 & 8.52   \\
HD 211706  & 80.1  &-106.1  & -12.6  & -63.11(0.17) & 2.99 & 8.55  \\
HD 214059  & 99.1  &-101.6  & -112.8 & -11.77(0.12) & 3.08 & 8.93  \\
C-40 15036  & 88.4  &-79.7   & -9.1   & -13.04(0.15) & 3.91 & 8.86  \\
HD 219180  & 59.6  &-89.9   &  2.0   & -29.41(0.20) & 3.66 & 8.32  \\
HD 220536  & 7.4   &-96.7   &  2.1   &   1.34(0.21) & 3.48 & 8.00   \\
HD 220993  &-22.4  &-96.7   & -18.9  &  51.30(0.18) & 3.44 & 8.13   \\
HD 224383  &74.5   &- 84.4  & -0.8   & -31.01(0.26) & 3.81 & 8.56   \\
HD   4308  &-51.5  &-110.4  &-29.5   &  95.69(0.16) & 2.87 & 8.41   \\
HD   6734  &-51.2  &-122.6  & 38.1   & -94.49(0.15) & 2.47 & 8.38   \\
HD   8638  & 26.1  &-86.3   &-77.7   &  84.67(0.27) & 3.89 & 8.04    \\
HD   9424  & 57.0  &-96.5   &-1.5    &  43.34(0.19) & 3.41 & 8.27   \\
HD  10576  & 55.9  &-92.6   &-19.2   &  54.39(0.19) & 3.57 & 8.27  \\
HD  10785  & 34.3  &-103.1  &13.7    &  -5.13(0.15) & 3.21 & 8.07  \\
HD  11306  & 26.7  &-97.3   &-19.1   &  64.88(0.22) & 3.45 & 8.03  \\
HD  11397  & -19.1 &-93.3   &-49.7   &  41.05(0.15) & 3.58 & 8.11  \\
HD  14282  & 74.6  &-90.4   & 41.3   &   0.19(0.26) & 3.58 & 8.53   \\
HD  16623  & -12.4 &-91.2   & 2.8    &  17.87(0.32) & 3.68 & 8.07   \\
BD-02 603  & -46.8 &-112.4  & -48.2  &   6.55(0.19) & 2.81 & 8.35   \\
HD  21543  & 57.6  &-91.5   &-19.9   &  64.22(0.24) & 3.60 & 8.29   \\
\enddata
\end{deluxetable}

\begin{deluxetable}{crrrrrrrrr}
\label{tb2}
\tabletypesize{\scriptsize}
\tablecaption{Sample stars.}
\tablewidth{0pt}
\tablehead
{
\colhead{Number} &\colhead{Name} & \colhead{V}   & \colhead{Spectral Type}
& \colhead{r(pc)} & \colhead{E(B-V)} & \colhead{$\rm M_{v}$} & \colhead{BC}
& \colhead{$\rm M_{\ast}$/$\rm M_{\odot}$} & \colhead {S/N}
}
\startdata
b1  & HD 143016 & 8.50  &  G3V    & 58.11 & 0.094 & 4.40 &  -0.274 & $0.85^{1}$ & 154 \\
b2  & HD 143102 & 7.88  &  G6IV/V & 70.77 & 0.027 & 3.55 &  -0.231 & $1.10^{1}$ & 127 \\
b3  & HD 148530 & 8.81  &  G9V    & 46.51 & 0.040 & 5.35 &  -0.307 & $0.90^{1}$ & 120 \\
b4  & HD 149256 & 8.42  &  G8IV   & 75.07 & 0.077 & 3.80 &  -0.311 & $0.80^{2}$ &  71 \\
b5  & HD 152391 & 6.65  &  G8V    & 16.94 & 0.033 & 5.40 &  -0.299 & $0.90^{1}$ & 133 \\
b6  & HDE326583 & 9.48  &  G0    & 143.88 & 0.101 & 3.38 &  -0.222 & $1.00^{1}$ & 121 \\
b7  & HD 175617 & 10.12 &  G5     & 74.74 & 0.100 & 5.44 &  -0.265 & $0.80^{1}$ & 124 \\
b8  & HD 178737 & 8.72  &  G2/G3V & 86.28 & 0.049 & 3.89 &  -0.274 & $0.90^{1}$ & 125 \\
b9  & HD 179764 & 9.01  &  G5     & 62.58 & 0.056 & 4.85 &  -0.250 & $0.90^{2}$ & 129 \\
b10 & HD 181234 & 8.59  &  G5     & 48.80 & 0.045 & 5.01 &  -0.311 & $0.90^{2}$ &  91 \\
b11 & HD 184846 & 8.95  &  G5V    & 81.83 & 0.037 & 4.27 &  -0.262 & $0.85^{1}$ & 118 \\
b12 & BD-176035 & 10.30 &  -      & 70.27 & 0.037 & 5.95 &  -0.538 & $0.85^{2}$ &  54 \\
b13 & HD 198245 & 8.94  &  G3V    & 59.59 & 0.033 & 4.96 &  -0.274 & $0.80^{1}$ & 151 \\
b14 & HD 201237 & 10.10 &  K2III  & 89.05 & 0.045 & 5.21 &  -0.538 & $0.95^{2}$ &  73 \\
b15 & HD 211276 & 8.72  &  G5     & 66.49 & 0.034 & 4.50 &  -0.372 & $0.85^{1}$ & 195 \\
b16 & HD 211532 & 9.31  &  G5     & 52.74 & 0.015  & 5.65 &  -0.316 & $0.80^{1}$ & 160 \\
b17 & HD 211706 & 8.90  &  G0     & 84.89 & 0.041  & 4.13 &  -0.206 & $1.00^{1}$ & 136 \\
b18 & HD 214059 & 8.26  &  G4V    & 80.45 & 0.021  & 3.67 &  -0.265 & $0.90^{1}$ & 142 \\
b19 & CD-4015036& 10.08 &  G5V    & 93.02 & 0.033  & 5.14 &  -0.312 & $0.90^{1}$ & 107 \\
b20 & HD 219180 & 9.81  &  G5V    & 80.39 & 0.032  & 5.18 &  -0.274 & $0.80^{1}$ & 132 \\
b21 & HD 220536 & 9.01  &  G1V    & 87.72 & 0.032  & 4.19 &  -0.217 & $0.95^{1}$ & 112 \\
b22 & HD 220993 & 9.34  &  G3V    & 92.59 & 0.032  & 4.41 &  -0.262 & $0.90^{1}$ & 118 \\
b23 & HD 224383 & 7.89  & G2V     & 47.66 & 0.032  & 4.40 &  -0.206 & $1.00^{1}$ & 126 \\
b24 & HD   4308 & 6.55  &  G5V    & 21.85 & 0.011  & 4.82 &  -0.265 & $0.90^{1}$ & 149 \\
b25 & HD   6734 & 6.44  &  K0IV   & 46.45 & 0.032  & 3.01 &  -0.363 & $1.05^{1}$ & 163 \\
b26 & HD   8638 & 8.29  &  G3V    & 40.40 & 0.032  & 5.16 &  -0.265 & $0.85^{1}$ & 153 \\
b27 & HD   9424 & 9.17  &  G8V    & 59.03 & 0.049  & 5.16 &  -0.312 & $0.90^{1}$ & 104 \\
b28 & HD  10576 & 8.51  &  G0/G1V & 85.76 & 0.062  & 3.65 &  -0.201 & $1.00^{1}$ & 145 \\
b29 & HD  10785 & 8.51  &  G1/G2V & 70.03 & 0.032  & 4.19 &  -0.211 & $0.95^{1}$ & 159 \\
b30 & HD  11306 & 9.24  &  G8V    & 75.02 & 0.057  & 4.69 &  -0.321 & $0.85^{1}$ & 144 \\
b31 & HD  11397 & 8.96  &  G6IV/V & 54.64 & 0.032  & 5.17 &  -0.282 & $0.80^{1}$ & 147 \\
b32 & HD  14282 & 8.39  &  G0     & 87.72 & 0.032  & 3.58 &  -0.226 & $0.95^{1}$ & 127 \\
b33 & HD  16623 & 8.76  &  G2V/VI & 65.88 & 0.032  & 4.57 &  -0.235 & $0.90^{1}$ & 108 \\
b34 & BD-02 603 & 9.03  &  G0    & 109.53 & 0.044  & 3.70 &  -0.253 & $0.90^{1}$ &  97 \\
b35 & HD  21543 & 8.23  &  G2V/VI & 45.50 & 0.020  & 4.88 &  -0.249 & $0.70^{1}$ & 135 \\
\enddata
\tablerefs{(1) Vandenberg (1985), (2) VandenBerg \& Laskarides (1987)}
\end{deluxetable}

\begin{deluxetable}{crrrrrrrrrrr}
\label{tb3}
\tablecolumns{10}
\tablewidth{0pc}
\tablecaption{Effective Temperatures}
\tablewidth{0pt}
\tablehead{
\colhead{Number} & \colhead{T$^{\rm Gen}$} & \colhead{T$^{\rm H\alpha}$} & 
\colhead{T$^{\rm Strom}$} & \colhead{T$^{\rm IRFM}$} & \colhead{Number} & \colhead{T$^{\rm Gen}$} & 
\colhead{T$^{\rm H\alpha}$} & \colhead{T$^{\rm Strom}$} & \colhead{T$^{\rm IRFM}$} 
}
\startdata
b1  &  5575 & 5575 & -            &	-    & b19 &  5340 & 5350 &        -   &    -  \\
b2  &  5432 & 5500 &   5397  &	-    & b20 &  5417 & 5400 &       -    &    -  \\
b3  &  5346 & 5350 &   5576  &	-    & b21 &  5829 & 5850 &  5868 &    -  \\
b4  &  5271 & 5350 & -            &	-    & b22 &  5599 & 5600 &       -    &      -	\\
b5  &  5335 & 5300 &   5366  &	-    & b23 &  5689 & 5800 &  5715 &  -	\\
b6  &  5629 & 5600 & -            &	-    & b24 &  5581 & 5600 &  5609 &  -	\\
b7  &  5456 & 5550 &   5451  & 5656$^{1}$& b25&  5021 & 5000 &  5000 &  -	\\
b8  &  5495 & 5575 & -            &	-    & b26 &  5469 & 5500 &  5445 &  -	\\
b9  &  5373 & 5450 & -            &	-    & b27 &  5332 & 5350 &  5358 &  -	\\
b10 &  5220 & 5350 &   5201  &	-    & b28 &  5883 & 5850 &  5915 &  -	\\
b11 &  5587 & 5600 &   5593  &	-    & b29 &  5773 & 5850 &  5782 &  -	\\
b12 &  4830 & 4750 &       -      &	-    & b30 &  5207 & 5200 &  5253 &  -	\\
b13 &  5588 & 5650 &   5606  &	-    & b31 &  5435 & 5400 &  5412 &  -	\\
b14 &  4886 & 4950 & 	      -   &	-    & b32 &  5790 & 5800 &       -    &  -	\\
b15 &  5372 & 5500 &   5404  &	-    & b33 &  5714 & 5700 &  5851 &  -	\\
b16 &  5295 & 5350 & 	       -  &	-    & b34 &  5476 & 5450 &       -    &  -	\\
b17 &  5830 & 5800 &   5865  &	-    & b35 &  5625 & 5650 &  5613 & 5568$^{1}$ \\
b18 &  5468 & 5550 &   5504  & 5553$^{2}$&    &       &      & 	       &	   \\
\enddata
\tablerefs{(1) Alonso et al. (1996), (2) Blackwell et al. (1990)}
\end{deluxetable}

\begin{deluxetable}{cccccccc}
\label{tb4}
\tabletypesize{\scriptsize}
\tablecaption{Effective temperatures check}
\tablewidth{0pt}
\tablehead{
\colhead{Number} &\colhead{T$^{\rm Gen}$} & \colhead{T$^{\rm H\alpha}$} & \colhead{T$^{\rm Edv}$} &
\colhead{T$^{\rm Kur}$} & \colhead{T$^{\rm H\beta}$} & \colhead{\rm T$^{\rm Strom}$}
}
\startdata
b01  &   5575  &   5575  &   5575  &   5500  &   5575  &    -     \\
b14  &   4886  &   4950  &   4750  &   4750  &   4950  &    -     \\
b15  &   5372  &   5500  &   5400  &   5500  &   5500  &   5404  \\
b19  &   5340  &   5350  &   5250  &   5250  &     -   &     -    \\
b22  &   5599  &   5600  &   5500  &   5500  &   5600  &     -    \\
b23  &   5689  &   5800  &   5700  &   5750  &   5800  &   5715 \\
b25  &   5021  &   5000  &   5000  &   5000  &   5000  &   5000 \\
b28  &   5883  &   5850  &   5850  &   5750  &   5850  &   5915 \\
b29  &   5773  &   5850  &   5750  &   5750  &   5850  &   5782 \\
b30  &   5207  &   5200  &   5200  &   5000  &   5200  &   5253  \\
\enddata
\end{deluxetable}

\begin{deluxetable}{crrrrrrrr}
\label{tb5}
\tabletypesize{\scriptsize}
\tablecolumns{8}
\tablewidth{0pc}
\tablecaption{\ion{Fe}{1} line list.}
\tablehead{
\colhead{Ion} & \colhead{$\lambda(\rm \AA)$} & \colhead{$\chi_{\rm ex}$} &
\colhead{log gf} & \colhead{Ion} & \colhead{$\lambda(\rm \AA)$} & \colhead{$\chi_{\rm ex}$} &
\colhead{log gf}
}
\startdata
\ion{Fe}{1} &  6056.005 &  4.73 &   -0.46 &  \ion{Fe}{1} &  6546.239 &  2.76 &  -1.54 \\
\ion{Fe}{1} &  6079.009 &  4.65 &   -1.13 &  \ion{Fe}{1} &  6569.215 &  4.73 &  -0.42 \\
\ion{Fe}{1} &  6082.711 &  2.22 &   -3.57 &  \ion{Fe}{1} &  6575.016 &  2.59 &  -2.82 \\
\ion{Fe}{1} &  6093.644 &  4.61 &   -1.51 &  \ion{Fe}{1} &  6593.870 &  2.43 &  -2.42 \\
\ion{Fe}{1} &  6096.665 &  3.98 &   -1.93 &  \ion{Fe}{1} &  6597.561 &  4.79 &  -1.06 \\
\ion{Fe}{1} &  6105.131 &  4.55 &   -2.06 &  \ion{Fe}{1} &  6608.026 &  2.28 &  -4.04 \\
\ion{Fe}{1} &  6151.623 &  2.18 &   -3.30 &  \ion{Fe}{1} &  6609.110 &  2.56 &  -2.69 \\
\ion{Fe}{1} &  6157.733 &  4.07 &   -1.25 &  \ion{Fe}{1} &  6627.545 &  4.55 &  -1.68 \\
\ion{Fe}{1} &  6173.336 &  2.22 &   -2.88 &  \ion{Fe}{1} &  6633.412 &  4.83 &  -1.49 \\
\ion{Fe}{1} &  6180.204 &  2.73 &   -2.78 &  \ion{Fe}{1} &  6633.750 &  4.56 &  -0.78 \\
\ion{Fe}{1} &  6187.990 &  3.94 &   -1.72 &  \ion{Fe}{1} &  6634.107 &  4.79 &  -1.43 \\
\ion{Fe}{1} &  6200.313 &  2.61 &   -2.44 &  \ion{Fe}{1} &  6646.934 &  2.61 &  -3.99 \\
\ion{Fe}{1} &  6213.430 &  2.22 &   -2.65 &  \ion{Fe}{1} &  6703.567 &  2.76 &  -3.15 \\
\ion{Fe}{1} &  6219.280 &  2.20 &   -2.43 &  \ion{Fe}{1} &  6704.476 &  4.22 &  -2.66 \\
\ion{Fe}{1} &  6220.794 &  3.88 &   -2.46 &  \ion{Fe}{1} &  6710.319 &  1.48 &  -4.87 \\
\ion{Fe}{1} &  6240.646 &  2.22 &   -3.39 &  \ion{Fe}{1} &  6713.745 &  4.79 &  -1.60 \\
\ion{Fe}{1} &  6254.258 &  2.28 &   -2.48 &  \ion{Fe}{1} &  6716.237 &  4.58 &  -1.93 \\
\ion{Fe}{1} &  6265.134 &  2.18 &   -2.55 &  \ion{Fe}{1} &  6725.357 &  4.10 &  -2.30 \\
\ion{Fe}{1} &  6270.225 &  2.86 &   -2.71 &  \ion{Fe}{1} &  6726.667 &  4.61 &  -1.00 \\
\ion{Fe}{1} &  6280.618 &  0.86 &   -4.39 &  \ion{Fe}{1} &  6733.151 &  4.64 &  -1.57 \\
\ion{Fe}{1} &  6297.793 &  2.22 &   -2.74 &  \ion{Fe}{1} &  6750.152 &  2.42 &  -2.60 \\
\ion{Fe}{1} &  6302.494 &  3.69 &   -0.91 &  \ion{Fe}{1} &  6752.707 &  464  &  -1.37 \\
\ion{Fe}{1} &  6311.500 &  2.83 &   -3.22 &  \ion{Fe}{1} &  6786.860 &  4.19 &  -2.06 \\
\ion{Fe}{1} &  6315.811 &  4.07 &   -1.71 &  \ion{Fe}{1} &  6793.259 &  4.07 &  -2.47 \\
\ion{Fe}{1} &  6322.685 &  2.59 &   -2.43 &  \ion{Fe}{1} &  6806.845 &  2.73 &  -3.21 \\
\ion{Fe}{1} &  6335.331 &  2.20 &   -2.18 &  \ion{Fe}{1} &  6810.263 &  4.61 &  -1.11 \\
\ion{Fe}{1} &  6336.824 &  3.69 &   -1.05 &  \ion{Fe}{1} &  6820.372 &  4.64 &  -1.31 \\
\ion{Fe}{1} &  6344.149 &  2.43 &   -2.92 &  \ion{Fe}{1} &  6828.591 &  4.64 &  -0.92 \\
\ion{Fe}{1} &  6355.029 &  2.84 &   -2.29 &  \ion{Fe}{1} &  6841.339 &  4.61 &  -0.75 \\
\ion{Fe}{1} &  6380.743 &  4.19 &   -1.40 &  \ion{Fe}{1} &  6842.686 &  4.64 &  -1.31 \\
\ion{Fe}{1} &  6419.950 &  4.73 &   -0.25 &  \ion{Fe}{1} &  6843.656 &  4.55 &  -0.93 \\
\ion{Fe}{1} &  6430.849 &  2.18 &   -2.01 &  \ion{Fe}{1} &  6855.162 &  4.56 &  -1.82 \\
\ion{Fe}{1} &  6475.624 &  2.56 &   -2.94 &  \ion{Fe}{1} &  6978.852 &  2.48 &  -2.50 \\
\ion{Fe}{1} &  6481.970 &  2.28 &   -2.98 &  \ion{Fe}{1} &  7038.223 &  4.22 &  -1.31 \\
\ion{Fe}{1} &  6495.742 &  4.83 &   -0.94 &  \ion{Fe}{1} &  7068.410 &  4.07 &  -1.38 \\
\ion{Fe}{1} &  6496.467 &  4.79 &   -0.57 &  \ion{Fe}{1} &  7090.383 &  4.23 &  -1.21 \\
\ion{Fe}{1} &  6498.939 &  0.96 &   -4.70 &  \ion{Fe}{1} &  7130.922 &  4.22 &  -0.78 \\
\ion{Fe}{1} &  6518.367 &  2.83 &   -2.75 &  \ion{Fe}{1} &  7132.986 &  4.07 &  -1.75 \\
\ion{Fe}{1} &  6533.929 &  4.56 &   -1.45 &  \ion{Fe}{1} &  7401.685 &  4.19 &  -1.69 \\
\enddata
\end{deluxetable}

\begin{deluxetable}{crrrrrrrr}
\label{tb6}
\tabletypesize{\scriptsize}
\tablecolumns{8}
\tablewidth{0pc}
\tablecaption{\ion{Fe}{2} line list.}
\tablewidth{0pt}
\tablehead{
\colhead{Ion} & \colhead{$\lambda(\rm \AA)$} & \colhead{$\chi_{\rm ex}$} & \colhead{log gf} &
\colhead{Ion} & \colhead{$\lambda(\rm \AA)$} & \colhead{$\chi_{\rm ex}$} & \colhead{log gf}
}
\startdata
\ion{Fe}{2} & 4508.29 & 2.85 &  -2.21 & \ion{Fe}{2} & 5325.55  & 3.22 & -2.60 \\
\ion{Fe}{2} & 4515.34 & 2.48 &  -2.48 & \ion{Fe}{2} & 5414.07  & 3.22 & -3.36 \\
\ion{Fe}{2} & 4520.22 & 2.81 &  -2.61 & \ion{Fe}{2} & 5425.26  & 3.20 & -3.79 \\
\ion{Fe}{2} & 4534.17 & 2.85 &  -3.48 & \ion{Fe}{2} & 5534.85  & 3.24 & -2.92 \\
\ion{Fe}{2} & 4555.89 & 2.83 &  -2.28 & \ion{Fe}{2} & 6084.105 & 3.20 & -4.06 \\
\ion{Fe}{2} & 4576.34 & 2.03 &  -3.04 & \ion{Fe}{2} & 6113.329 & 3.22 & -4.35 \\
\ion{Fe}{2} & 4582.83 & 2.84 &  -3.09 & \ion{Fe}{2} & 6233.498 & 4.48 & -2.84 \\
\ion{Fe}{2} & 4620.52 & 2.81 &  -3.29 & \ion{Fe}{2} & 6239.948 & 3.89 & -3.57 \\
\ion{Fe}{2} & 4656.98 & 2.89 &  -3.63 & \ion{Fe}{2} & 6247.562 & 3.89 & -2.68 \\
\ion{Fe}{2} & 4670.18 & 2.58 &  -4.10 & \ion{Fe}{2} & 6369.463 & 2.89 & -1.90 \\
\ion{Fe}{2} & 4731.45 & 2.89 &  -3.37 & \ion{Fe}{2} & 6383.715 & 4.55 & -2.44 \\
\ion{Fe}{2} & 4833.20 & 2.66 &  -4.78 & \ion{Fe}{2} & 6385.458 & 4.55 & -2.83 \\
\ion{Fe}{2} & 4923.93 & 2.89 &  -1.32 & \ion{Fe}{2} & 6416.928 & 3.89 & -2.74 \\
\ion{Fe}{2} & 5000.74 & 2.77 &  -4.75 & \ion{Fe}{2} & 6432.683 & 2.89 & -3.55 \\
\ion{Fe}{2} & 5132.67 & 2.81 &  -4.17 & \ion{Fe}{2} & 6456.391 & 3.90 & -2.43 \\
\ion{Fe}{2} & 5197.58 & 3.23 &  -2.10 & \ion{Fe}{2} & 6493.060 & 4.58 & -3.00 \\
\ion{Fe}{2} & 5234.62 & 3.22 &  -2.05 & \ion{Fe}{2} & 6516.083 & 2.89 & -3.65 \\
\ion{Fe}{2} & 5284.11 & 2.89 &  -3.20 & \ion{Fe}{2} & 7067.460 & 3.8  & -3.85 \\
\enddata
\end{deluxetable}

\begin{deluxetable}{crrrrrrrrrrrrrrrrrrr}
\label{tb7}
\tabletypesize{\scriptsize}
\tablecolumns{14}
\tablecaption{Measured equivalent widths of \ion{Fe}{1} lines: b1 to b17}
\tablewidth{0pt}
\tablehead{
\colhead{$\lambda$} & \colhead{b1} & \colhead{b2} & \colhead{b3} & \colhead{b4} & \colhead{b5}
& \colhead{b6} & \colhead{b7} & \colhead{b8} & \colhead{b9} & \colhead{b10} & \colhead{b11}
& \colhead{b12} &  \colhead{b13} & \colhead{b14} & \colhead{b15} & \colhead{b16} & \colhead{b17}
}
\startdata
6056.005 &  52 &  87 &  79 &  89 &  85 &  55 &  58 &  54 &  86 & 103 &  63 & 95  & 47 & 77  & 53 & 52 &70 \\
6079.009 &  31 &  60 &  47 &  68 &  54 &   - &  33 &  38 &  49 &  76 &  37 & 62  & 23 & 62  & 44 & 36 &51 \\
6082.711 &  27 &  56 &  52 &  67 &  51 &   - &  29 &  34 &  51 &  66 &   - & 73  & 15 & 65  & 28 & 30 &33 \\
6093.644 &  23 &  44 &  40 &  55 &  41 &  24 &  18 &  27 &  44 &  58 &  24 & 51  & 15 & 44  & 20 & 20 &38 \\
6096.665 &  25 &  53 &  50 &  64 &  51 &  25 &  26 &  32 &  52 &  68 &  38 & 73  & 20 & 62  & 30 & 25 &40 \\
6105.131 &  7  &  24 &  16 &  41 &  19 &   - &   6 &   - &  67 &  34 &  10 & 31  & -  & 18  & 8  & 14 &15 \\
6151.623 &  39 &  68 &  67 &  79 &  65 &  43 &  42 &  47 &  59 &  82 &  45 & 86  & 29 & 79  & 37 & 43 &49 \\
6157.733 &  43 &  78 &  69 &  91 &  70 &  50 &  50 &  60 &  81 &  90 &  55 & 93  & 40 & 83  & 48 & 48 &66 \\
6173.336 &  54 &  87 &  82 & 100 &  85 &  51 &  60 &   - &  83 &  99 &  66 & 110 & 48 & 98  & 52 & 57 &13 \\
6180.204 &  37 &  76 &  69 &  87 &  72 &  39 &  43 &  64 &  72 & 102 &  48 & 91  & 30 & 80  & 29 & 46 &60 \\
6187.990 &  28 &  69 &  56 &  78 &  63 &  29 &   - &   - &   - &  72 &  40 & 75  & 24 & 65  & 33 & 31 &49 \\
6200.313 &  48 &  94 &  82 & 110 &  89 &  52 &  65 &  64 &  90 & 116 &  65 & 105 & 48 & 93  & 18 & 57 &70 \\
6213.430 &  64 & 113 &  90 & 111 & 106 &  63 &  73 &  75 &  96 & 118 &  76 & 115 & 58 & 102 & 64 & 60 &80 \\
6220.794 &  -  &  29 &  23 &  43 &   - &   - &   - &   - &   - &  50 &  85 & 41  & 7  & 36  & 18 & 14 &-  \\
6232.641 &  -  &  98 &  99 & 113 & 107 &  63 &  71 &  70 & 100 &   - &  75 & 116 & 51 & 115 & 72 & 82 &50 \\
6240.646 &  32 &  71 &  45 &  79 &  63 &  34 &  41 &  43 &  64 &  85 &  44 & 80  & 31 & 68  & 37 & 40 &50 \\
6254.258 & 100 & 136 & 126 & 153 & 137 &   - & 100 &   - &   - & 157 & 118 & 156 & 91 & 143 & 90 & 10 &124\\
6265.134 &  71 &  -  & 108 & 122 & 108 &  77 &  79 &  78 & 100 &   - &  83 &   - & -  & 137 & 67 & 76 &90 \\
6270.225 &  49 &  -  &  63 &  86 &  65 &  46 &  42 &  46 &  69 &   - &  49 &   - & -  & 79  & 41 & 42 &60 \\
6280.618 &  56 &  -  &  90 & 103 & 103 &   - &   - &  67 &   - &   - &  65 &   - & -  & 109 & 63 & 66 &70 \\
6297.793 &  84 & 106 &  93 & 111 & 119 &  73 &  70 &  72 &  90 & 110 &  72 & 123 & 54 & 113 & 56 & 63 &73 \\
6302.494 &  64 & 101 &  97 & 110 & 109 &  72 &  71 &  77 & 105 & 140 &  79 & 142 & 58 & 124 & 63 & 48 &85 \\
6311.500 &  16 &  44 &  44 &  58 &  44 &  18 &  21 &  22 &  52 &  59 &  25 &  72 & 13 & 53  & 25 & 23 &28 \\
6315.811 &  28 &  59 &  49 &  68 &  54 &  32 &  36 &  31 &  59 &  69 &  35 &  71 & 23 & 56  & 26 & 51 &48 \\
6322.685 &  59 &  96 &  96 & 115 &  96 &  67 &   - &  75 &  91 & 116 &  74 & 121 & 53 & 114 & 59 & 65 &76 \\
6335.331 &  86 & 118 & 115 & 135 & 126 &  89 &  87 &  91 & 119 & 142 &  94 & 174 & 79 & 154 & 78 & 88 &100\\
6336.824 &  83 & 115 & 123 & 134 & 144 &  91 &   - &   - &   - & 165 & 101 & 192 & 78 & 174 & 83 & 82 &104\\
6344.149 &  -  &  73 &  72 &  -  &  81 &  48 &  51 &  51 &  71 &  95 &  52 &  93 & 39 & 95  & 50 & 52 &-  \\
6355.029 &  58 &  99 &  97 & 128 & 101 &  55 &  62 &  66 &  98 & 129 &  70 & 122 & 47 & 118 & 55 & 45 &82 \\
6380.743 &  37 &  66 &  56 &  78 &  64 &  41 &  39 &  45 &  65 &  81 &  52 &  72 & 28 & 55  & 40 & 35 &49 \\
6419.950 &  60 &  99 &  90 & 112 & 102 &  66 &  71 &  70 &  99 & 124 &  77 & 103 & 57 & 97  & -  & 15 &89 \\
6430.849 &  90 & 125 & 139 & 155 &   - & 100 & 101 & 109 & 137 & 170 & 108 & 225 & 88 & 193 & 89 & 10 &100\\
6475.624 &  37 &  71 &  72 &  90 &  77 &  43 &  45 &  50 &  69 &  98 &  53 & 100 & 41 & 84  & 38 & 48 &59 \\
6481.970 &  49 &  81 &  80 &  99 &  90 &  54 &  55 &  62 &  78 & 100 &  60 & 109 & 42 & 98  & 51 & 54 &-  \\
6495.742 &  25 &  51 &  47 &  66 &  48 &   - &  29 &   - &  51 &  76 &   - &  76 & 23 & 62  & 27 & 28 &43 \\
6496.467 &  49 &  74 &  78 &  90 &  74 &  44 &  45 &  44 &  79 & 109 &  53 &  88 & 36 & 85  & 49 & 46 &66 \\
6498.939 &  30 &  60 &  66 &  85 &  65 &  39 &  41 &   - &  65 &  83 &  43 &  99 & 30 & 90  & 31 & 59 &52 \\
6518.367 &  40 &  72 &  74 &  88 &  71 &  46 &  47 &  54 &  71 &   - &   - &   - & -  & 90  & 45 & 44 &62 \\
6533.929 &  20 &  51 &  50 &  70 &   - &  27 &  23 &  28 &  49 &  69 &  33 &  61 & 20 & 55  & 25 & 25 &46 \\
6546.239 &  76 & 115 & 118 & 134 & 133 &   - & 104 &  96 & 123 & 155 & 100 & 163 & 82 & 158 & 90 & 95 &98 \\
6569.215 &  54 &  83 &  81 &  96 &  90 &  52 &  61 &  60 &  85 & 107 &  62 & 102 & 49 & 94  & 62 & 55 &74 \\
6575.016 &  41 &  80 &  70 &  94 &  79 &  45 &  53 &  57 &  77 &  97 &  56 & 110 & 40 & 96  & 52 & 54 &57 \\
6593.870 &  67 & 104 &  95 & 113 & 109 &  66 &  76 &   - & 103 & 125 & 103 & 123 & 82 & 111 & 64 & 71 &81 \\
6597.561 &  28 &  55 &  51 &  65 &  56 &  29 &  33 &  33 &  51 &  75 &  34 &  63 & 22 & 52  & 27 & 32 &46 \\
6608.026 &   9 &  30 &  27 &  44 &  29 &   - &  15 &  13 &  31 &  49 &  18 &  53 & 8  & 42  & 13 & 18 &20 \\
6609.110 &  51 &   - &  77 & 108 &  86 &  54 &  61 &  56 &  82 & 111 &  60 & 106 & 41 & 84  & 52 & 51 &66 \\
6627.545 &  17 &  44 &  37 &  54 &  37 &  18 &  18 &  20 &  38 &  55 &  23 &  54 & 15 & 45  & 15 & 18 &27 \\
6633.412 &  14 &  38 &  35 &  63 &  37 &  16 &  17 &  55 &  38 &  61 &  23 &  64 & 14 & 52  & 19 & 19 &35 \\
6633.750 &  48 &  79 &  77 &  93 &  77 &  51 &  50 &  32 &  75 &  98 &  58 &  94 & 41 & 86  & 47 & 51 &68 \\
6634.107 &  24 &   - &  46 &  -  &  42 &  25 &  21 &   - &  48 &   - &  34 &  73 & 17 & 55  & 39 & 24 &52 \\
6646.932 &  -  &  19 &  20 &  33 &  16 &   - &   - &   - &   - &   - &   - &   - & -  & 33  & 6  & 7  &-  \\
6703.567 &  23 &  54 &  54 &  72 &  49 &  28 &  34 &  32 &  50 &  73 &  34 &  66 & 19 & 70  & 28 & 33 &41 \\
6710.319 &  8  &  28 &  30 &  53 &  26 &   - &   - &   - &   - &  51 &  18 &  66 & 6  & 50  & 11 & 14 &16 \\
6713.745 &  15 &  28 &  35 &  47 &  29 &  18 &   - &   - &   - &  43 &  18 &   - & -  & 57  & 13 & -  &26 \\
6716.237 &  10 &  27 &  25 &  39 &  27 &   - &   - &   - &   - &  42 &  16 &  43 & 7  & 28  & 16 & 10 &20 \\
6725.357 &  15 &  28 &  28 &  41 &  25 &  12 &   - &   - &   - &  75 &  16 &  46 & -  & 37  & 13 & 10 &17 \\
6726.667 &  28 &  59 &  59 &  69 &  59 &  38 &  33 &  38 &  56 &  75 &  43 &  76 & 28 & 68  & 33 & 33 &55 \\
6733.151 &  14 &  39 &  38 &  55 &  32 &  17 &  15 &  21 &  37 &  53 &  22 &  51 & 13 & 39  & 17 & 17 &29 \\
6750.152 &  56 &  93 &  86 & 108 &  89 &  60 &  64 &  67 &  89 & 112 &  70 & 102 & 52 & 98  & 60 & 61 &73 \\
6752.707 &  -  &  50 &  43 &  73 &  45 &  25 &  21 &  26 &  42 &  72 &  31 &  67 & 18 & 43  & 27 & 20 &36 \\
6786.860 &  -  &  40 &  35 &  49 &  33 &  15 &  15 &  17 &  34 &  50 &  21 &  43 & 11 & 36  & -  & 15 &26 \\
6793.259 &  -  &  21 &  19 &  36 &  20 &   - &  8  &  10 &  19 &  32 &  13 &  28 & -  & 25  & 9  & 8  &12 \\
6806.845 &  20 &  52 &  47 &  66 &  46 &  23 &  28 &  29 &  50 &  70 &  34 &  68 & 16 & 60  & 24 & 27 &35 \\
6810.263 &  35 &  63 &  58 &  77 &  60 &  36 &  35 &  41 &  62 &  81 &  40 &  68 & 24 & 66  & 34 & 34 &52 \\
6820.372 &  27 &  57 &  51 &  68 &  -  &  32 &  25 &  31 &  50 &  71 &  35 &   7 & 22 & 58  & 27 & 28 &41 \\
6828.591 &  42 &  69 &  67 &  82 &  68 &  42 &  41 &  46 &  67 &  89 &  51 &  84 & 33 & 75  & 42 & 41 &65 \\
6841.339 &  53 &  76 &  83 &  96 &  85 &  58 &  53 &  60 &   - & 109 &  64 &   - & 40 & 88  & 53 & 48 &-  \\
6842.866 &  24 &  47 &  30 &  55 &  48 &  28 &  27 &  33 &  50 &  65 &  33 &  64 & 17 & 57  & 26 & -  &-  \\
6943.656 &   - &  70 &  72 &  91 &  74 &  44 &  49 &  55 &  73 &  88 &  57 &  86 & 35 & 80  & 42 & 55 &-  \\
6855.162 &  58 &  84 &  79 & 103 &  85 &  60 &  58 &  59 &  85 & 103 &  65 & 102 & -  & 95  & 53 & 54 &68 \\
6978.852 &  64 &  91 &  84 & 110 &  98 &  61 &  67 &  68 &  92 & 113 &  76 &  95 & 51 & 99  & 77 & 65 &82 \\
7038.223 &  48 &  75 &  77 & 106 &   - &  50 &  48 &  59 &   - &  95 &  57 &   - & 40 & 104 & 45 & 46 &64 \\
7068.410 &  48 &  84 &  80 & 118 &  81 &  38 &  47 &  57 &  85 &  76 &  62 & 113 & 38 & 100 & 38 & 30 &70 \\
7090.383 &  51 &  83 &  81 &  98 &  81 &  57 &  51 &  56 &  84 & 110 &  62 & 111 & 39 & 98  & 45 & 53 &69 \\
7130.922 &  68 & 109 &  99 & 138 & 111 &  67 &  73 &  74 & 104 & 136 &  80 & 139 & 63 & 130 & 60 & 69 &96 \\
7132.986 &  32 &  53 &  50 &  65 &  53 &  32 &  36 &  33 &  52 &  69 &  36 &   - & 23 & 56  & 26 & 30 &49 \\
7401.685 &  34 &  56 &  50 &  71 &  54 &   - &  -  &   - &   - &   - &  36 &   - & -  & 54  & 25 & 34 &52 \\
\enddata
\end{deluxetable}

\begin{deluxetable}{crrrrrrrrrrrrrrrrrrrr}
\label{tb8}
\tabletypesize{\scriptsize}
\tablecolumns{14}
\tablecaption{Measured equivalent widths of \ion{Fe}{1} lines: b18 to b35}
\tablewidth{0pt}
\tablehead{
\colhead{$\lambda$} & \colhead{b18} & \colhead{b19} & \colhead{b20} & \colhead{b21} &
\colhead{b22} & \colhead{b23} & \colhead{b24} & \colhead{b25} &
\colhead{b26} & \colhead{b27} & \colhead{b28} & \colhead{b29} &  \colhead{b30} & \colhead{b31}
& \colhead{b32} & \colhead{b33} & \colhead{b34} & \colhead{b35}
}
\startdata
6056.005 & 66 & 75 & 42 & 57 & 63 & 79 & 59 & 64  & 61 & 86  & 73  & 67 & 47 & 66 & 57 & -  & -  & 51\\
6079.009 & 36 & 47 & 28 & 41 & 38 & 46 & 34 & 39  & 32 & 54  & -   & 40 & 30 & 35 & 28 & 23 & 22 & 29\\
6082.711 & 32 & 45 & -  & 30 & 34 & 36 & -  & 48  & 30 & 51  & 30  & 25 & 21 & 34 & 18 & 20 & 24 & - \\
6093.644 & 26 & 38 & 19 & 26 & 22 & 37 & 25 & 28  & 22 & 45  & 31  & 27 & 15 & 21 & -  & -  & -  & 14\\
6096.665 & 35 & 49 & 26 & 30 & 32 & 39 & 30 & 37  & 26 & 56  & 37  & 30 & 22 & 33 & 22 & 17 & 23 & 22\\
6105.131 & 10 & 23 & -  & -  & 11 & -  & 10 & 12  & -  & 24  & 12  & 9  & -  & -  & -  & -  & -  & - \\
6151.623 & 49 & 64 & 41 & 37 & 46 & 49 & 40 & 62  & 42 & 68  & 48  & 44 & 34 & 51 & 33 & 31 & 42 & 33\\
6157.733 & 55 & 76 & 41 & 50 & 55 & 65 & 50 & 63  & 54 & 77  & 62  & 56 & 43 & 53 & 43 & 44 & 42 & 41\\
6173.336 & 63 & 78 & 52 & 57 & 64 & 70 & 56 & 81  & 62 & 87  & 64  & 61 & 52 & 63 & 46 & 50 & 51 & 49\\
6180.204 & 49 & 69 & 34 & 41 & 51 & 60 & 46 & 65  & 47 & 73  & 49  & 49 & 36 & 55 & 36 & 34 & 36 & 36\\
6187.990 & 39 & 54 & 27 & 36 & 42 & 48 & 36 & 49  & 38 & 59  & 43  & 41 & 27 & 40 & 30 & 29 & 28 & 28\\
6200.313 & 65 & 77 & -  & 61 & 63 & 75 & 64 & 82  & 66 & 85  & 74  & 68 & 59 & 71 & 56 & 55 & 46 & 52\\
6213.430 & 76 & 87 & 59 & 66 & 75 & 88 & 71 & 93  & 78 & 93  & 81  & 77 & 69 & 86 & 66 & 66 & 56 & 66\\
6220.794 & -  & 95 & -  & 80 & -  & -  & -  &  -  & -  & 103 & -   &  - &  - &  - & -  & 72 & 61 & 72\\
6232.641 & -  & -  & -  & -  & -  & -  & -  &  -  & -  & 66  & -   &  - &  - &  - & -  & 30 & 33 & 33\\
6240.646 & 45 & 54 & 30 & 33 & 41 & 51 & 40 & 62  & 44 & 135 & 44  & 42 & 33 & 49 & 30 & -  & -  & - \\
6254.258 & 117& 137& 80 & 10 & 118& 11 & -  & 121 & -  & 106 & -   & 91 & -  & 111& 76 & 65 & 72 & 67\\
6265.134 & 85 & 103& 70 & 76 & 82 & 88 & 76 & 99  & 80 & 70  & 86  & 80 & 68 & 90 & 73 & 34 & 34 & 34\\
6270.225 & 52 & 66 & 39 & 36 & 54 & 54 & 45 & 62  & 44 & 92  & 50  & 45 & 35 & 50 & 30 & 44 & -  & 47\\
6280.618 & 69 & 99 & 88 & 51 & 62 & -  & -  & 89  & 67 & 96  & -   & -  & -  & 78 & -  & 54 & 63 & 56\\
6297.793 & 68 & 92 & 62 & 65 & 72 & 78 & 67 & 87  & 69 & 101 & -   & 69 & 70 & 100& 60 & 81 & -  & 62\\
6302.494 & 75 & 99 & 87 & 85 & 76 & 10 & 73 & 77  & 75 & 43  & 82  & 76 & 68 & 84 & -  & 11 & -  & 13\\
6311.500 & 28 & 44 & -  & 22 & 27 & 30 & 18 & 39  & 20 & 57  & 23  & 22 & 18 & 34 & -  & 22 & 26 & 21\\
6315.811 & 35 & 51 & 28 & 31 & 67 & 45 & 33 & 42  & 31 & 99  & 42  & 36 & 23 & 34 & 24 & 53 & 58 & 60\\
6322.685 & 72 & 92 & 59 & 62 & 78 & 78 & 67 & 83  & 72 & 121 & 75  & 70 & 58 & 76 & 59 & 78 & 78 & 78\\
6335.331 & 91 & 118& 84 & 86 & 95 & 99 & 87 & 106 & 93 & 131 & 97  & 90 & 82 & 103& 82 & 76 & -  & - \\
6336.824 & 93 & 136& 88 & 92 & 102& 10 & 92 & 101 & -  & -   & 103 & 95 & -  & 109& 77 & -  & -  & - \\
6344.149 & -  & -  & 47 & 49 & 53 & -  & -  & 71  & -  & 101 & -   & -  & -  & -  & -  & 52 & 48 & 50\\
6355.029 & 69 & 95 & 54 & 62 & 71 & 80 & 67 & 91  & 69 & 58  & 72  & 66 & 56 & 72 & 50 & 31 & -  & 36\\
6380.743 & 46 & 55 & 31 & 38 & 44 & 51 & 43 & 53  & 45 & 96  & 50  & 49 & 37 & 38 & 41 & 63 & 51 & 63\\
6419.950 & 72 & 86 & 55 & 67 & 78 & 87 & 72 & 74  & 75 & 136 & 88  & 78 & 64 & 74 & 67 & 93 &    & 93\\
6430.849 & 108& 131& 98 & 10 & 110& 11 & 10 & 124 & 10 & 72  & 105 & 105& 95 & -  & 95 & -  & 45 & 36\\
6475.624 & 52 & 72 & 35 & 53 & 49 & 66 & 44 & 69  & 49 & 84  & 54  & 53 & 36 & 53 & 41 & 44 & 52 & 47\\
6481.970 & 61 & 78 & 52 & 49 & 61 & 67 & 56 & 76  & 58 & 57  & 65  & 59 & 50 & 71 & 48 & 33 &    & 20\\
6495.742 & 41 & 45 & -  & 37 & 153& 38 & 27 & 34  & -  & 81  & 36  & 32 & 19 & -  & -  & 43 & 49 & 44\\
6496.467 & 53 & 68 & 46 & 53 & 57 & 74 & 51 & 56  & 52 &     & 66  & 56 & 43 & 56 & 39 & 26 & 35 & 29\\
6498.939 & 45 & 66 & 42 & 28 & 43 & 46 & 36 & 69  & 41 & 74  & 41  & 35 & 31 & 53 & 25 &    & 46 & 43\\
6518.367 & 54 & 73 & 46 & 43 & 55 & 60 & 49 & 65  & 49 & -   & 54  & 54 & 42 & -  & 40 & 23 & -  & - \\
6533.929 & 33 & 46 & 82 & 32 & -  & 40 & 27 & 34  & 27 & 121 & 36  & 31 & -  & 18 & -  & 79 & 78 & 83\\
6546.239 & 100& 119& -  & 85 & 96 & 10 & 98 & 122 & 97 & 84  & 102 & 92 & 89 & 113& 88 & 54 & 44 & 51\\
6569.215 & 61 & 76 & 48 & 56 & 68 & 72 & 61 & 68  & 63 & 74  & 73  & 66 & 51 & 68 & 55 & 42 & 45 & 40\\
6575.016 & 60 & 76 & 41 & 46 & 53 & 61 & 52 & 74  & 56 & 98  & 56  & 68 & 44 & 62 & 54 & 63 & 54 & 63\\
6593.870 & 78 & 88 & 83 & 68 & 78 & 84 & 75 & 95  & 79 & 54  & 83  & 79 & 67 & 88 & 66 & 23 & -  & 21\\
6597.561 & 37 & 46 & 29 & 33 & 39 & 48 & 34 & 38  & 34 & 29  & 42  & 35 & 27 & 34 & 24 & -  & -  & 12\\
6608.026 & 18 & 31 & 11 & 10 & 18 & 21 & 13 & 29  & 13 & 81  & 12  & 13 & -  & 23 & -  & 45 & 43 & 49\\
6609.110 & 56 & 76 & 45 & 51 & 59 & 70 & 55 & 75  & 58 & 44  & 62  & 61 & 48 & 65 & 47 & -  & -  & 15\\
6627.545 & 21 & 35 & 20 & 22 & 23 & 32 & 19 & 26  & 20 & 40  & 27  & 22 & -  & 21 & 15 & -  & -  & 16\\
6633.412 & 24 & 37 & 18 & 19 & 22 & 31 & 18 & 26  & 20 & 77  & 28  & 23 & 13 & 16 & 17 & 44 & 44 & 48\\
6633.750 & 58 & 77 & 51 & 54 & 59 & 65 & 54 & 60  & 56 & -   & 62  & 58 & 45 & 55 & 47 & -  & -  & - \\
6634.107 & 32 & 44 & 25 & 25 & 34 & -  & -  & 40  & -  & -   & -   & -  & -  & 26 &  - & -  &- &  - \\
6646.932 & -  & 15 & -  & -  & -  & -  & -  & 15  & -  & 55  & 10  & 8  & -  & -  & 7  & 22 & -  & 22\\
6703.567 & 31 & 47 & 26 & 28 & 32 & 38 & 30 & 48  & 29 & -   & 33  & 31 & 21 & 33 & 23 & -  & -  & - \\
6710.319 & 15 & 26 & -  & -  & 14 & -  & 12 & 32  & 12 & -   & 11  & 11 & -  & 18 & -  & -  & -  & - \\
6713.745 & 19 & 49 & -  & 23 & 20 & 27 & -  & 18  & 14 & 29  & 22  & 18 & -  & 11 & -  & -  & -  & - \\
6716.237 & -  & 19 & 8  & -  & 12 & 18 & -  & 18  & 10 & 29  & -   & 11 & -  & -  & -  & -  & -  & - \\
6725.357 & 16 & 28 & -  & 13 & -  & 20 & 11 & 20  & 13 & 62  & 16  & 15 & -  & -  & -  & 23 & -  & 31\\
6726.667 & 40 & 58 & 33 & 39 & 44 & 48 & 35 & 42  & 37 & 33  & 46  & 41 & 31 & 38 & -  & -  & -  & - \\
6733.151 & 23 & 34 & 15 & 20 & 21 & 28 & 21 & 26  & 17 & 88  & 22  & 20 & -  & -  & 17 & 56 &  - & 57\\
6750.152 & 67 & 81 & 50 & 60 & 66 & 73 & 64 & 85  & 67 & 48  & 72  & 67 & 59 & 72 &    & 16 & 19 & 19\\
6752.707 & 29 & 42 & 20 & 30 & 30 & 33 & 28 & 35  & 27 & 34  & 36  & 30 & 22 & 26 & 19 &  - &  - & - \\
6786.860 & 16 & 31 & 13 & 18 & 24 & 24 & 14 & 25  & 16 & -   & 19  & 22 & -  & -  & -  & -  & -  & - \\
6793.259 & -  & 16 & -  & -  & -  & -  & 9  & 14  & -  & 53  & -   & -  & -  & -  & -  & 21 & -  & 21\\
6806.845 & 32 & 44 & 23 & 26 & 29 & 38 & 27 & 47  & 28 & 61  & 32  & 26 & 22 & 33 & 20 & 33 & -  & 30\\
6810.263 & 40 & 54 & 30 & 40 & 43 & 54 & 42 & 44  & 40 & 56  & 47  & 43 & 31 & 43 & 34 & 21 & -  & 24\\
6820.372 &  27 &  57 &  51 &  68 &  -  &  32 &  25 &  31 &  50 &  71 &  35 &   7 & 22 & 58  & 27 & 28 &41 \\
6828.591 &  42 &  69 &  67 &  82 &  68 &  42 &  41 &  46 &  67 &  89 &  51 &  84 & 33 & 75  & 42 & 41 &65 \\
6841.339 &  53 &  76 &  83 &  96 &  85 &  58 &  53 &  60 &   - & 109 &  64 &   - & 40 & 88  & 53 & 48 &-  \\
6842.866 &  24 &  47 &  30 &  55 &  48 &  28 &  27 &  33 &  50 &  65 &  33 &  64 & 17 & 57  & 26 & -  &-  \\
6943.656 &   - &  70 &  72 &  91 &  74 &  44 &  49 &  55 &  73 &  88 &  57 &  86 & 35 & 80  & 42 & 55 &-  \\
6855.162 &  58 &  84 &  79 & 103 &  85 &  60 &  58 &  59 &  85 & 103 &  65 & 102 & -  & 95  & 53 & 54 &68 \\
6978.852 &  64 &  91 &  84 & 110 &  98 &  61 &  67 &  68 &  92 & 113 &  76 &  95 & 51 & 99  & 77 & 65 &82 \\
7038.223 &  48 &  75 &  77 & 106 &   - &  50 &  48 &  59 &   - &  95 &  57 &   - & 40 & 104 & 45 & 46 &64 \\
7068.410 &  48 &  84 &  80 & 118 &  81 &  38 &  47 &  57 &  85 &  76 &  62 & 113 & 38 & 100 & 38 & 30 &70 \\
7090.383 &  51 &  83 &  81 &  98 &  81 &  57 &  51 &  56 &  84 & 110 &  62 & 111 & 39 & 98  & 45 & 53 &69 \\
7130.922 &  68 & 109 &  99 & 138 & 111 &  67 &  73 &  74 & 104 & 136 &  80 & 139 & 63 & 130 & 60 & 69 &96 \\
7132.986 &  32 &  53 &  50 &  65 &  53 &  32 &  36 &  33 &  52 &  69 &  36 &   - & 23 & 56  & 26 & 30 &49 \\
7401.685 &  34 &  56 &  50 &  71 &  54 &   - &  -  &   - &   - &   - &  36 &   - & -  & 54  & 25 & 34 &52 \\
\enddata
\end{deluxetable}

\begin{deluxetable}{crrrrrrrrrrrrrrrrrrrr}
\label{tb9}
\tabletypesize{\scriptsize}
\tablecolumns{14}
\tablewidth{0pc}
\tablecaption{Measured equivalent widths of \ion{Fe}{2} lines: b1 to b17}
\tablewidth{0pt}
\tablehead{
\colhead{$\lambda$} & \colhead{b1} & \colhead{b2} & \colhead{b3} & \colhead{b4} & \colhead{b5}
& \colhead{b6} & \colhead{b7} & \colhead{b8} & \colhead{b9} & \colhead{b10} & \colhead{b11}
& \colhead{b12} &  \colhead{b13} & \colhead{b14} & \colhead{b15} & \colhead{b16} & \colhead{b17}
}
\startdata
4508.29 & 80 & 104 & 77  & 100 & 81  & 78  & 65  & 79  & 84  & 85 & 79 & 65 & 68 & 61 & 71 & 46 & 100\\
4515.34 & 73 & 102 & 83  & 108 & 87  & 70  & 58  & 78  & 84  & 115& 78 & 95 & 59 & -  & 69 & 51 & 96 \\
4520.22 & 75 & 93  & 74  & 95  & 79  & 73  & 62  & 77  & 77  & 85 & 76 & 72 & 58 & 67 & 68 & 49 & 91 \\
4534.17 & 48 & 62  & 44  & 63  & 45  & 42  & 28  & 41  & 46  & 56 & 44 & -  & 28 & 32 & -  & 17 & 62 \\
4555.89 & 83 & 103 & 82  & 107 & 89  & 77  & 65  & 83  & 86  & 87 & 81 & 63 & 66 & -  & 68 & 58 & -  \\
4576.34 & -  & -   & 53  & -   & 56  & -   & -   & 57  & -   & 64 & 54 & -  & 41 & 39 & 49 & -  & 71 \\
4582.83 & 46 &     &     & 90  & 53  & 41  & 37  & 47  & 61  & -  & 45 & -  & 31 & -  & 38 & -  & 64 \\
4620.52 & 45 & 76  & 49  & 75  & 47  & 46  & 35  & 53  & 55  & 73 & 48 & -  & 31 & 32 & 42 & -  & 60 \\
4656.98 & 27 & 48  & 91  & 46  & 24  & -   & -   & 27  & 30  & 34 & 22 & -  & -  & -  & 23 & 11 & 40 \\
4670.18 & 22 & 48  & 29  & 48  & 28  & 19  & 12  & 24  & 37  & 34 & 24 & -  & 12 & 17 & -  & -  & 38 \\
4731.45 & -  & -   & 78  & -   & 81  & -   & -   & -   & -   & 91 & 71 & -  & -  & -  & -  & -  & -  \\
4833.20 & 9  &     & 10  & 24  & -   & -   & -   & 11  & 12  & -  & 8  & -  & -  & -  & -  & -  & 17 \\
4923.93 & 134& 181 & 135 & 171 & 153 & 133 & 111 & 136 & 140 & 148& 137& 100& 11 & 101& 123& 99 & 171\\
5000.74 & -  & -   & 7   & 15  & 5   & -   & -   & 7   & -   & 8  & 3  & -  & -  & -  & -  & -  & 11 \\
5132.67 & -  & -   & -   & -   & 15  & 21  & -   & 22  & 27  & -  & 17 & -  & -  & -  & 16 & -  & 32 \\
5197.58 & 71 &     & 72  & 97  & 75  & 66  & 60  & 74  & 79  & 90 &    & 46 & 57 & 42 & 67 & 34 & 89 \\
5234.62 & 76 & 99  & 72  & 95  & 77  & 78  & 58  & 81  & 77  & 80 &    & 51 & 61 & 47 & 61 & -  & 84 \\
5284.11 & 52 & -   & 62  & 81  & 64  & 51  & 41  & 57  & 65  & 70 & 55 &    & 36 & -  & -  & 29 & 71 \\
5325.55 & -  & -   & 32  & -   & 33  & -   & -   & -   & -   & 39 & 31 & -  & -  & -  & -  & -  & -  \\
5414.07 & 24 & 44  & 19  & 42  & 20  & -   & -   & -   & 24  & 29 & 21 & -  & -  & -  & 15 & 8  & 40 \\
5425.26 &    &     & 36  &     & 37  & -   & -   & -   & -   & 42 & 35 & -  & -  & -  & -  & -  & -  \\
5534.85 & 53 &     & 47  & 63  & 47  & 54  & 35  & 57  & 53  & 55 & 49 & 35 & 36 & 31 & -  & -  & 72 \\
6084.10 & -  & 32  & 17  & 35  & 17  & 18  & 10  & 18  & 19  & 28 & -  & 14 & 14 & -  & -  & -  & -  \\
6113.33 & 6  & 28  & 8   & 22  & 11  & -   & -   & 13  & -   & 18 & -  & 11 & -  & -  & -  & -  & 23 \\
6196.68 & -  & -   & -   & -   & -   & -   & -   & -   & -   & 51 & -  & -  & -  & 4  & -  & -  & -  \\
6233.50 & -  &     & 12  & -   & -   & -   & -   & -   & -   & 11 & -  & -  & -  & -  & -  & -  & -  \\
6239.95 & -  & 26  & -   & -   & 16  & -   & 4   & -   & -   & -  & 11 & -  & -  & -  & -  & -  & -  \\
6247.56 & 43 & 71  & 43  & -   & -   & 45  & 32  & 46  & 45  & 48 & 48 & 22 & 32 & 19 & -  & 26 & 66 \\
6369.46 & 13 & 28  & 13  & 30  & -   & -   & -   & -   & 18  & 26 & 17 & -  & 8  & 5  & -  & 6  & -  \\
6383.71 & 6  & 17  & -   & 19  & 11  & -   & -   & -   & 9   & 19 & -  & -  & -  & 10 & -  & -  & 18 \\
6385.46 & -  & 15  & -   & -   & -   & -   & -   & -   & 16  & -  & -  & -  & -  & -  & -  & -  & -  \\
6416.93 & -  & -   & -   & -   & -   & -   & -   & 31  & 40  & 50 & 35 & 10 & 19 & 32 & -  & 26 & -  \\
6432.68 & 38 & 51  & 36  & 56  & 33  & 37  & 23  & -   & 37  & 33 & 37 & 39 & 22 & 20 & -  & 15 & 59 \\
6456.39 & 55 & 77  & 49  & 74  & 56  & 57  & 39  & 58  & 56  & 61 & 64 & 36 & 41 & 32 & -  & 24 & 83 \\
6493.06 & 5  & -   & -   & -   & -   & -   & -   & -   & -   & -  & -  & -  & -  & -  & -  & -  & 7  \\
6516.08 & 46 & 85  & 38  & 81  & 45  & 45  & 31  & 50  & 52  & 61 & 50 & 29 & 29 & 27 & -  & 24 & 77 \\
7067.46 & 7  & 17  & -   & -   & 15  & -   & -   & -   & 11  & 31 & -  & 22 & 5  & 15 & -  & -  & 14 \\
\enddata
\end{deluxetable}

\begin{deluxetable}{crrrrrrrrrrrrrrrrrrrr}
\label{tb10}
\tabletypesize{\scriptsize}
\tablecolumns{14}
\tablewidth{0pc}
\tablecaption{Measured equivalent widths of \ion{Fe}{2} lines: b18 to b35}
\tablewidth{0pt}
\tablehead{
\colhead{$\lambda$} & \colhead{b18} & \colhead{b19} & \colhead{b20} & \colhead{b21} &
\colhead{b22} & \colhead{b23} & \colhead{b24} & \colhead{b25} &
\colhead{b26} & \colhead{b27} & \colhead{b28} & \colhead{b29} &  \colhead{b30} & \colhead{b31}
& \colhead{b32} & \colhead{b33} & \colhead{b34} & \colhead{b35}
}
\startdata
4508.29 & 84 & 77 & 65 & - & 81 & 86 & 79 & 73 & 73 & 83 &  99 & 88 &  65 & 69 & 88 & 77 & 63 & 67\\
4515.34 & 71 & 79 & 56 & - & 75 & 88 & 68 & 75 & 63 & 80 &  97 & 85 &  59 & 59 & 83 & 73 & -  & 62\\
4520.22 & 73 & 74 & -  & - & 76 & 84 & 71 & 71 & 65 & 79 &  92 & 83 &  61 & 60 & 80 & 70 & 57 & 69\\
4534.17 & 43 & 42 & -  & - & 39 & 53 & 41 & 38 & 36 & 46 &  61 & 50 &  -  & -  & 50 & 37 & -  & 36\\
4555.89 & 81 & 77 & 58 & - & 80 & 92 & 77 & 75 & 70 & 88 &  99 & 90 &  62 & 66 & 85 & 75 & 64 & 71\\
4576.34 & -  & -  & -  & - & -  & -  & -  & -  & 49 & -  &  75 & 65 &  -  & -  & -  & -  & -  & 48\\
4582.83 & 49 & 45 & 27 & - & 43 & 64 & 44 & 52 & 39 & -  &  68 & 58 &  32 & 34 & 52 & 41 & 28 & 44\\
4620.52 & 50 & 44 & 29 & - & 51 & 56 & 44 & 42 & 38 & 51 &  64 & 45 &  30 & 35 & 50 & 40 & 30 & 38\\
4656.98 & 30 & 20 & -  & - & 21 & 36 & 25 & 24 & 18 & 28 &  44 & 34 &  14 & 15 & 32 & 20 & 15 & 20\\
4670.18 & 24 & 25 & 11 & - & 22 & 32 & 18 & 22 & 16 & 26 &  36 & 30 &  11 & 16 & 25 & 17 & 17 & 18\\
4731.45 & -  & -  & -  & - & -  & -  & -  & -  & 64 & -  &  -  & -  &  -  & -  & -  & -  & -  & 59\\
4833.20 & -  & -  & -  & - & 7  & 12 & 9  & -  & -  & -  &  -  & -  &  -  & -  & 9  & -  & -  & - \\
4923.93 & 137& 132& 101& - & 137& 150& 130& 119& 12 & 141&  164& -  &  106& 117& 149& 127& 118& 12\\
5000.74 & -  & -  & -  &   & -  & -  & -  & -  & -  & -  &  -  & -  &  -  & -  & 11 & -  & -  & - \\
5132.67 & 20 & 18 & 9  & - & 19 & 28 & 17 & -  & -  & -  &  29 & -  &  -  & -  & 22 & -  & -  & 10\\
5197.58 & 72 & 64 & 39 & - & 72 & 86 & 71 & 71 & 64 & 72 &  99 & 88 &  65 & 59 & 84 & 69 & 50 & 65\\
5234.62 & 78 & 58 & 57 & - & 72 & 88 & -  & 67 & 65 & 78 &  95 & 86 &  53 & 63 & 86 & 69 & -  & 66\\
5284.11 & -  & 47 & 41 & - & 55 & 69 & 54 & 52 & 46 & -  &  77 &  - &  -  & 40 & 62 & 46 & -  & 45\\
5325.55 & -  & -  & -  & - & -  & -  & -  & -  & 22 & -  &  -  &  - &  -  & -  & -  & -  & -  & 23\\
5414.07 & -  & -  & -  & - & -  & -  & -  & -  & 12 & -  &  -  & 29 &  -  & -  & -  & 25 & -  & 14\\
5425.26 & 36 & -  & -  & - & -  & -  & -  & -  & 25 & -  &  -  &  - &  -  & -  & -  & -  & -  & 27\\
5534.85 & 52 & 48 & 340& - & 48 & 59 & 45 & 44 & 42 & 50 &  70 & 61 &  -  & 36 & 58 & -  & 38 & 41\\
6084.10 & 22 & 20 & -  & 28& 20 & -  & -  & 14 & 11 & 18 &  34 & 22 &  -  & -  & 24 & 14 & -  & - \\
6113.33 & 13 & 9  & 7  & 10& -  & -  & -  & 10 & -  & -  &  21 & 18 &  -  & -  & -  & -  & -  & 7 \\
6196.68 & -  & -  & -  & - & -  & -  & -  & -  & -  & -  &  -  &  - &  -  & -  & -  & -  & -  & - \\
6233.50 & -  & -  & -  & - & -  & -  & -  & -  & -  & -  &  -  &  - &  -  & -  & -  & -  & -  & - \\
6239.95 & -  & -  & -  & 13& 9  & -  & -  & -  & -  & -  &  23 & 15 &  -  & -  & 13 & -  & -  & - \\
6247.56 & 45 & 38 & 25 & 57& 43 & -  & -  & 35 & 32 & 42 &  69 & 59 &  29 & 33 & 58 & 40 & 24 & 37\\
6369.46 & 14 & -  & -  & 20& 16 & -  & -  & 11 & 11 & 19 &  27 & 20 &  -  & 8  & 16 & 14 & -  & - \\
6383.71 & 10 & 7  & -  & 10& -  & -  & -  & 5  & -  & -  &  14 & 10 &  -  & -  & -  & 5  & -  & - \\
6385.46 & -  & 4  & -  & 7 & -  & -  & -  & -  & -  & 19 &  -  &  - &  -  & -  & -  & -  & -  & 7 \\
6416.93 & 33 & -  & -  & - & -  & -  & -  & -  & -  & -  &  -  &  - &  -  & -  & -  & -  & 14 & - \\
6432.68 & 36 & 26 & -  & 46& 39 & -  & -  & 31 & 28 & 38 &  57 &  - &  20 & 19 & -  & 27 & -  & 31\\
6456.39 & 55 & 53 & 29 & 65& 61 & -  & -  & 48 & 42 & 57 &  74 &  64&  35 & 47 & 66 & -  & 35 & 45\\
6493.06 & -  & -  & -  & - & -  & -  & -  & 17 & -  & -  &  -  &  - &  -  & -  & -  & 5  & -  & - \\
6516.08 & 52 & 50 & 31 & 62& 53 & -  & -  & 43 & 37 & 54 &  66 &  57&  25 & 36 & 56 & 42 & 32 & 42\\
7067.46 & 13 & 17 & 42 & - & -  & -  & -  & 7  & -  & -  &  -  &  - &  -  & -  & -  & -  & -  & - \\
\enddata
\end{deluxetable}

\begin{deluxetable}{cccccccc}
\label{tb11}
\tablecolumns{8}
\tablewidth{0pc}
\tablecaption{Physical Parameters}
\tablehead{
\colhead{Reference} & \colhead{log g$^{\rm Hip}$} & \colhead{$\sigma$(log g$^{\rm Hip}$)} &
\colhead{log g$^{\rm Spec}$} & \colhead{[Fe/H]$^{\rm Gen}$} & \colhead{[Fe/H]$^{\rm Spec}$} &
\colhead{$\xi(\rm kms^{-1})$}
}
\startdata
b1  &  4.06  &  0.06   & 3.8 & -0.47 & -0.50 & 1.0 \\
b2  &  3.80  &  0.05   & 3.7 & 0.01  & 0.10  & 0.9 \\
b3  &  4.38  &  0.06   & 4.3 & 0.11  & 0.00  & 0.5 \\
b4  &  3.68  &  0.07   & 3.6 & 0.27  & 0.26  & 1.1 \\
b5  &  4.40  &  0.05   & 3.9 & -0.08 & -0.12 & 0.9 \\
b6  &  3.76  &  0.20   & 3.7 & -0.31 & -0.50 & 0.6 \\
b7  &  4.42  &  0.09   & 4.7 & -0.28 & -0.48 & 0.5 \\
b8  &  3.86  &  0.06   & 4.0 & -0.48 & -0.33 & 0.6 \\
b9  &  4.21  &  0.06   & 4.2 & 0.16  & 0.05  & 0.5 \\
b10 &  4.20  &  0.06   & 4.1 & 0.30  & 0.38  & 0.8  \\
b11 &  4.02  &  0.07   & 4.0 & -0.16 & -0.25 & 0.8  \\
b12 &  4.35  &  0.10   & 3.8 & 0.60  & 0.05  & 1.0 \\
b13 &  4.26  &  0.07   & 4.3 & -0.54 & -0.65 & 0.5 \\
b14 &  4.10  &  0.11   & 4.3 & 0.50  & -0.05 & 0.5 \\
b15 &  4.00  &  0.07   & 4.0 & -0.40 & -0.55 & 0.5 \\
b16 &  4.43  &  0.07   & 4.7 & -0.27 & -0.70 & 0.5 \\
b17 &  4.13  &  0.07   & 3.7 & 0.09  & -0.05 & 1.0 \\
b18 &  3.76  &  0.07   & 3.8 & -0.37 & -0.33 & 0.6 \\
b19 &  4.29  &  0.09   & 4.1 & 0.07  & -0.10 & 0.5 \\
b20 &  4.30  &  0.08   & 4.4 & -0.45 & -0.70 & 0.5  \\
b21 &  4.10  &  0.07   & 3.9 & -0.17 & -0.22 & 1.0  \\
b22 &  4.10  &  0.08   & 4.0 & -0.18 & -0.30 & 0.7  \\
b23 &  4.10  &  0.06   & 4.1 & 0.00  & -0.02 & 1.0   \\
b24 &  4.26  &  0.05   & 4.0 & -0.34 & -0.40 & 0.7  \\
b25 &  3.36  &  0.05   & 3.1 & -0.33 & -0.53 & 0.8  \\
b26 &  4.33  &  0.06   & 4.1 & -0.27 & -0.50 & 0.9  \\
b27 &  4.30  &  0.06   & 4.0 & 0.12  & 0.00  & 0.8   \\
b28 &  3.95  &  0.05   & 3.6 & 0.11  & -0.12 & 1.2  \\
b29 &  4.11  &  0.06   & 4.2 & -0.11 & -0.25 & 1.0   \\
b30 &  4.04  &  0.06   & 4.3 & -0.46 & -0.60 & 0.6  \\
b31 &  4.30  &  0.07   & 4.0 & -0.47 & -0.70 & 0.6  \\
b32 &  3.87  &  0.07   & 3.7 & -0.37 & -0.40 & 1.0  \\
b33 &  4.21  &  0.06   & 4.0 & -0.42 & -0.60 & 1.0  \\
b34 &  3.77  &  0.06   & 3.9 & -0.19 & -0.75 & 0.8  \\
b35 &  4.31  &  0.08   & 4.1 & -0.47 & -0.55 & 0.5  \\
\enddata
\end{deluxetable}

\begin{deluxetable}{cccccc}
\label{tb12}
\tablecolumns{5}
\tablewidth{0pc}
\tablecaption{Oxygen Abundances}
\tablehead{ \colhead{Name} & \colhead{[O/Fe]} & \colhead{Name} & \colhead{[O/Fe]}}
\startdata
HD 143016 & +0.23  & HD 214059  & - \\
HD 143102 & 0.00   & CD-40 1503  & - \\
HD 148530 & +0.33  & HD 219180  & - \\
HD 149256 & +0.08  & HD 220536  & 0.00 \\
HD 152391 &  0.00  & HD 220993  & +0.23 \\
HDE 326583 & +0.23  & HD 224383  & +0.18 \\
HD 175617 &   -    & HD   4308  & +0.18 \\
HD 178737 & +0.28  & HD   6734  & +0.28 \\
HD 179764 & +0.13  & HD   8638  & +0.23 \\
HD 181234 & -0.07  & HD   9424  & 0.00 \\
HD 184846 & +0.23  & HD  10576  & 0.00 \\
BD-17 6035 & 0.00   & HD  10785  & - \\
HD 198245 & -      & HD  11306  & +0.33 \\
HD 201237 & +0.08  & HD  11397  & +0.23 \\
HD 211276 & -      & HD  14282  &  -  \\
HD 211532 & +0.43  & HD  16623  & +0.33 \\
HD 211706 & -      & BD-02 603  & - \\
HD  21543 & +0.33  &    &  \\ 
\enddata
\end{deluxetable}

\clearpage

\begin{figure}
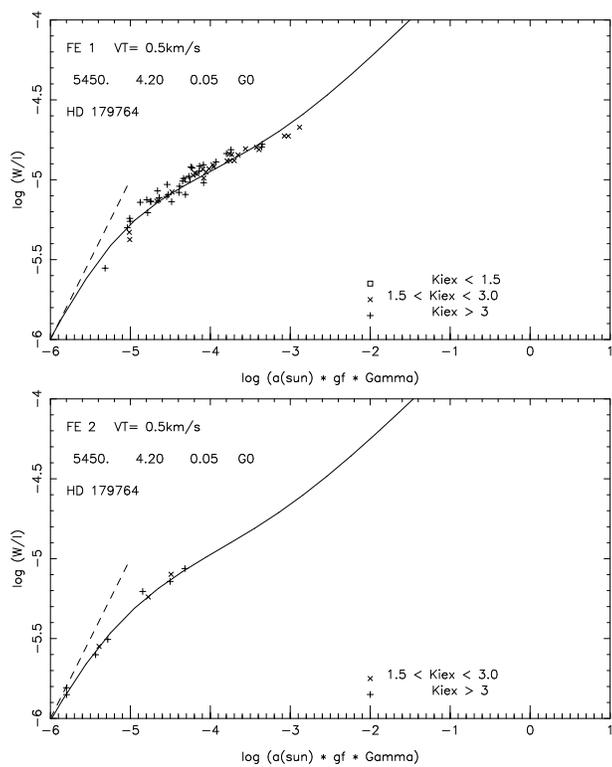

\label{fg2}\rotatebox{270}{\resizebox{!}{8cm}
{\includegraphics{PompeiaL1_fig2.eps}     
\includegraphics{PompeiaL1_fig3.eps}}}
\caption{Curves of growth for Fe I and Fe II lines of HD 179764 (symbols represent the lower
potentials of the lines).}
\end{figure}

\clearpage
\begin{figure}
\label{fg3}    {\resizebox{!}{6cm}
{\includegraphics{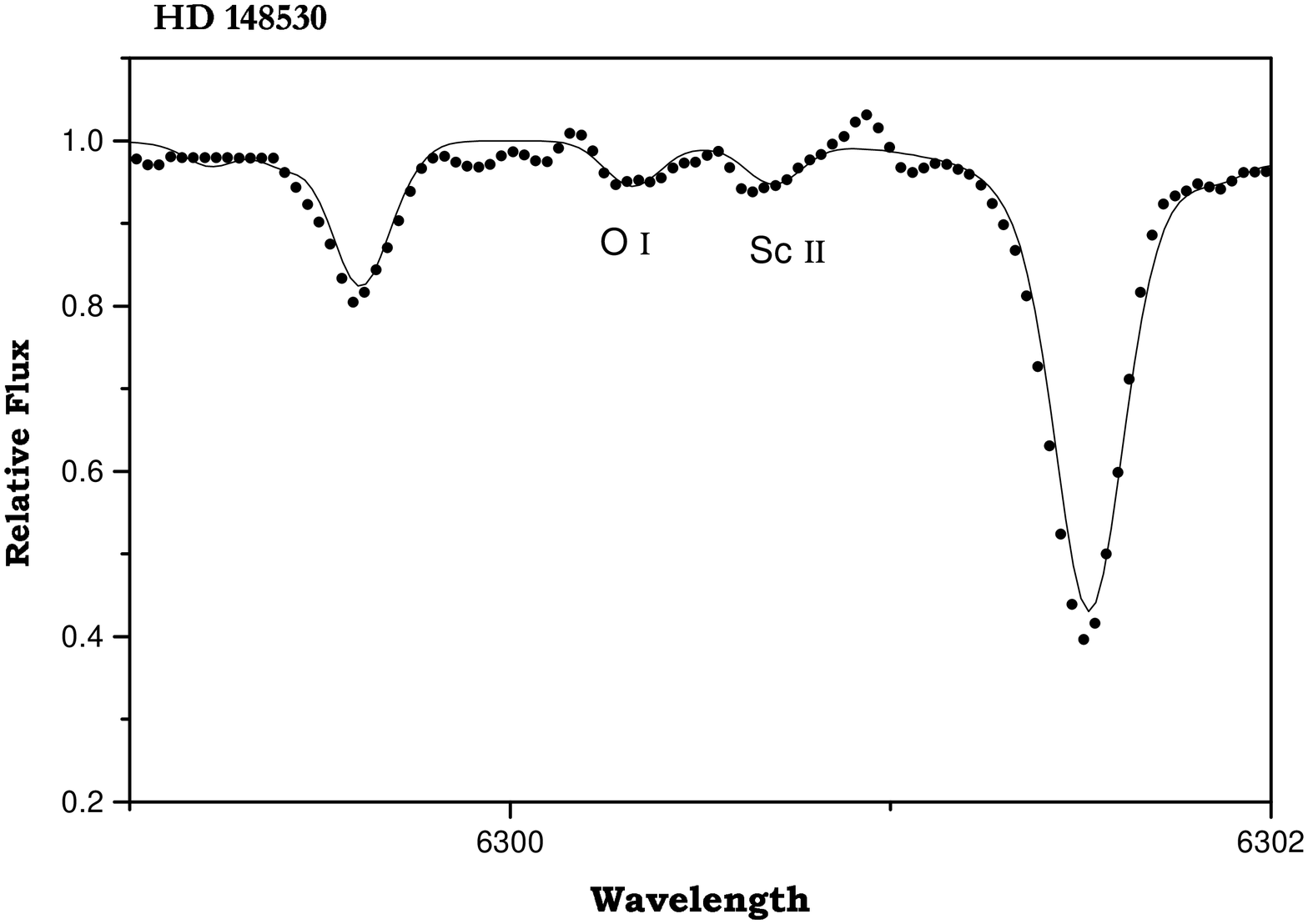}  \includegraphics{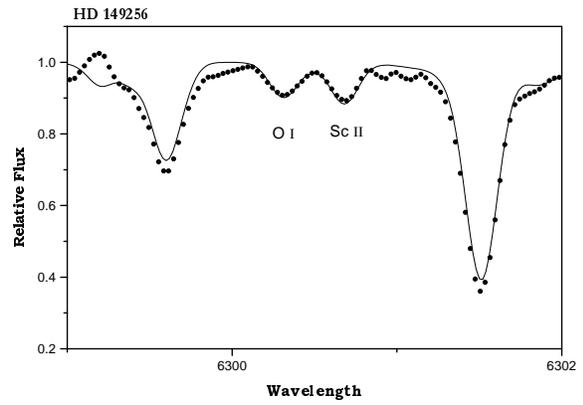}}}
\caption{Spectrum synthesis of the [OI] line: observed (dots) and synthetic (solid lines) for
HD 148530 and HD 149256.}
\end{figure}

\clearpage

\begin{figure}
\label{fg4}
\plotone{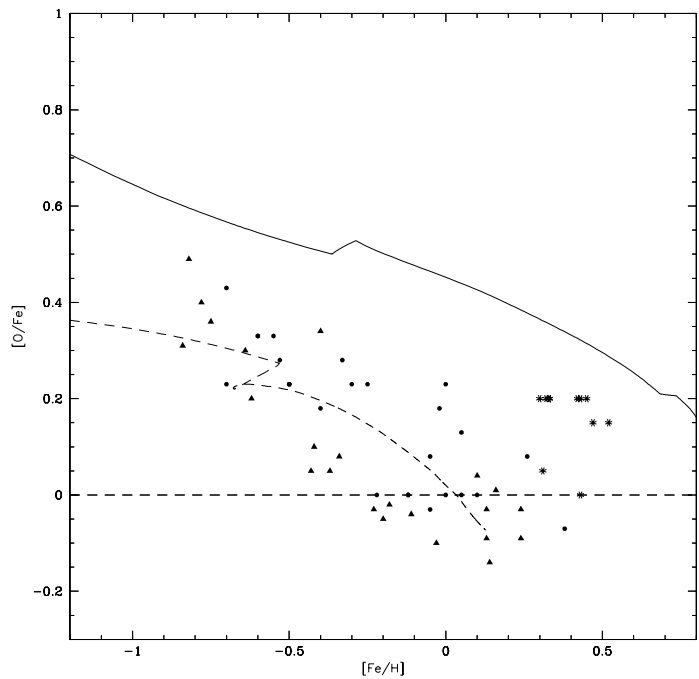}
\caption{Oxygen to iron ratio for three samples: our data (circles), SMR stars of Barbuy \& 
Grenon (1990) (stars) and a sample of disk stars (triangles) from Nissen \& Edvardsson (1992). 
The lines are theoretical curves for bulge (solid line) and solar vicinity (dashed line), 
from Matteucci et al. (1999) and Chiappini et al. (2001), respectively.}
\end{figure}

\clearpage


\end{document}